\documentclass[a4paper,11pt]{article}
\pdfoutput=1
\usepackage{jcappub_published}
\usepackage[british]{babel}
\usepackage{latexsym,amsmath,amssymb}
\usepackage{graphicx}
\usepackage{booktabs}
\usepackage{slashed}
\usepackage{bm}
\usepackage{epsfig}
\usepackage{dcolumn}

\newcommand{\exclude}[1]{}
\newcommand{\nn}{\nonumber}
\newcommand{\beq}{\begin{eqnarray}}
\newcommand{\eeq}{\end{eqnarray}}
\newcommand{\be}{\begin{eqnarray}}
\newcommand{\ee}{\end{eqnarray}}
\newcommand{\bea}{\begin{eqnarray}}
\newcommand{\eea}{\end{eqnarray}}

\newcommand{\rar}{\rightarrow}

\def\dd{ \,\text{d} }

\def\+{\dagger}
\def\la{\langle}
\def\ra{\rangle}
\def\<{\langle}
\def\>{\rangle}

\newcommand{\cH}{{\cal H}}
\newcommand{\cA}{{\cal A}}
\newcommand{\cP}{{\cal P}}
\newcommand{\cL}{{\cal L}}
\newcommand{\cC}{{\cal C}}

\newcommand{\Mpc}{\text{Mpc}}
\newcommand{\Gauss}{\text{Gauss}}
\newcommand{\mpl}{M_\text{Pl}}
\newcommand{\V}[1]{{\boldsymbol{#1}}}

\newcommand{\Vk}{\V{k}}

\newcommand{\Vq}{\V{q}}

\newcommand{\VB}{\V{B}}
\newcommand{\VE}{\V{E}}
\newcommand{\VJ}{\V{J}}
\newcommand{\VBtilde}{\widetilde{\VB}}
\newcommand{\VEtilde}{\widetilde{\VE}}
\newcommand{\VJtilde}{\widetilde{\VJ}}
\newcommand{\Veps}{\boldsymbol{\varepsilon}}

\newcommand{\mn}{{\mu\nu}}  
\newcommand{\ab}{{\alpha\beta}}  
\newcommand{\veps}{\varepsilon}  
\newcommand{\D}{{\tilde\nabla}}  
\newcommand{\cK}{{\cal K}}  
\renewcommand{\(}{\left(}
\renewcommand{\)}{\right)}
\renewcommand{\[}{\left[}
\renewcommand{\]}{\right]}
\newcommand{\order}[1]{{\cal O}(#1)}

\title{Cosmological Ohm's law and dynamics of non-minimal electromagnetism}

\author[a, b]{Lukas Hollenstein,}
\author[b, c]{Rajeev Kumar Jain}
\author[d, e]{and Federico R.~Urban}

\affiliation[a]{Institut de Physique Th{\'e}orique, CEA-Saclay, \\
Orme des Merisiers bat.\ 774, PC 136, 91191 Gif-sur-Yvette Cedex, France}

\affiliation[b]{D\'epartement de Physique Th\'eorique and Centre for Astroparticle Physics, Universit\'e de Gen\`eve, \\
24 Quai Ernest Ansermet, 1211 Gen\`eve 4, Switzerland}

\affiliation[c]{CP$^3$-Origins, Centre for Cosmology and Particle Physics Phenomenology, University of Southern Denmark, \\
Campusvej 55, 5230 Odense M, Denmark}

\affiliation[d]{Service de Physique Th\'eorique, Universit\'e Libre de Bruxelles, \\
CP225, Boulevard du Triomphe, B-1050 Brussels, Belgium}

\affiliation[e]{Institute for Theoretical Astrophysics, University of Oslo, \\
P.O.\ Box 1029 Blindern, N-0315 Oslo, Norway}

\emailAdd{lukas.hollenstein@cea.fr}
\emailAdd{jain@cp3.dias.sdu.dk}
\emailAdd{furban@ulb.ac.be}

\abstract{The origin of large-scale magnetic fields in cosmic structures and the intergalactic medium is still poorly understood. We explore the effects of non-minimal couplings of electromagnetism on the cosmological evolution of currents and magnetic fields. In this context, we revisit the mildly non-linear plasma dynamics around recombination that are known to generate weak magnetic fields. We use the covariant approach to obtain a fully general and non-linear evolution equation for the plasma currents and derive a generalised Ohm law valid on large scales as well as in the presence of non-minimal couplings to cosmological (pseudo-)scalar fields. Due to the sizeable conductivity of the plasma and the stringent observational bounds on such couplings, we conclude that modifications of the standard (adiabatic) evolution of magnetic fields are severely limited in these scenarios. Even at scales well beyond a Mpc, any departure from flux freezing behaviour is inhibited.}

\arxivnumber{1208.6547}

\begin{document}
\maketitle

\section{Introduction} \label{sec:intro}

Our Universe is apparently magnetised on virtually all different length-scales probed by astronomical observations. The strength of magnetic fields in galaxies and galaxy clusters is of the order of $\mu\Gauss$~\cite{Kronberg:1993vk,Han:2002ns,Govoni:2004as,Clarke:2000bz}, and a lower bound of the order of $10^{-15}\, \Gauss$ on coherent magnetic fields in the intergalactic medium has also been reported~\cite{Tavecchio:2010mk,Ando:2010rb,Neronov:1900zz,Dolag:2010ni,Essey:2010nd}. The origin of these large-scale magnetic fields still remains unclear, particularly at coherence lengths stretching beyond a Mpc, which is the subject of the present work\footnote{In this context, theoretical models operating during QCD or Electroweak phase transitions (recent representative works are~\cite{Saveliev:2012ea,Tevzadze:2012kk,Jedamzik:2010cy,Caprini:2009pr,Urban:2009sw}) will not be useful as the magnetic power spectrum they produce is cut off by causality at much smaller scales.}. Speculative mechanisms, active at all stages of the evolution of the Universe, have been proposed which often require the breaking of the conformal invariance of electromagnetism (EM). For recent reviews see refs.~\cite{Subramanian:2009fu,Kandus:2010nw,Widrow:2011hs,Ryu:2011hu}.

Besides these exotic possibilities, it is known that weak magnetic fields are generated around photon decoupling. In the pre-recombination era, the vorticities of electrons and protons evolve slightly differently yielding a net circular current that generates magnetic fields, an effect known as the Harrison mechanism~\cite{Harrison:1970,Harrison:1973zz}. However, no magnetic fields are produced at first order in cosmological perturbations, even in the presence of active sources of vector perturbations~\cite{Hollenstein:2007kg}. At second order in both perturbations and the tight coupling approximation of the photon-baryon interactions, the mechanism does generate magnetic fields of about $10^{-29}\,\Gauss$ on Mpc scales~\cite{Berezhiani:2003ik,Matarrese:2004kq,Gopal:2004ut,Ichiki:2007hu,Kobayashi:2007wd,Takahashi:2007ds,Maeda:2008dv,Fenu:2010kh,Maeda:2011uq,Giovannini:2011tj}.

This result provides a motivation to take a closer look at the evolution of EM fields during the late Universe. Clearly, in standard Maxwell theory the high conductivity of the cosmic plasma causes the electric fields to be screened rapidly and the magnetic fields to freeze into the plasma and evolve adiabatically, i.e.\ $B\propto a^{-2}$. However, one can expect a different behaviour in case Maxwell theory is modified on large scales. Specific scenarios have already been explored \cite{Lee:2001hj, Urban:2009sw, BeltranJimenez:2010uh}, but no systematic investigation has been carried out so far.

In this paper, we consistently examine whether the seed fields from the recombination era could be boosted at late times due to non-minimal (effective) interactions of EM. In particular, we allow for different extensions of EM to embrace couplings with a scalar or pseudo-scalar field, for instance, dark energy (DE) or a cosmological axion. In this context, we review the hydrodynamical evolution of the primordial plasma from electron-positron annihilation until today by carefully taking into account the effects of a possible non-minimal theory. To this end we employ the 1+3 covariant approach to cosmology that is fully non-linear, not restricted to a Friedmann-Lema\^{\i}tre-Robertson-Walker (FLRW) background, and captures all relativistic effects. Only after linearising, for illustration, it will be convenient to switch to a description using 3-vectors w.r.t.\ a comoving basis in a FLRW background.

First of all, in section~\ref{sec:ohm} we provide a brief review of the plasma processes occurring from the pre-recombination era to today by looking at the evolution of currents, relevant on cosmological scales at second order. Then, we derive a generalised Ohm's law that is also valid in the context of non-minimally coupled EM. By inspecting the different timescales involved in the current evolution we briefly discuss its validity on very large (sub- and super-horizon) cosmological scales. This section should serve as a reference for cosmological plasma conductivity formulae and physics.
In section~\ref{sec:modified}, we introduce the generic form of the Lagrangian allowing for different non-minimal couplings and discuss their effects on the magnetic field evolution at large scales. Finally, in section~\ref{TheEnd}, we summarise our results and discuss the conclusions. Appendix~\ref{app:conv} lists our conventions, and appendix~\ref{app:covariant} contains the details of the 1+3 covariant approach for the description of the fluid and current evolution in the plasma.

\section{Generalised Ohm's law for non-minimal electromagnetism}  \label{sec:ohm}

Circular currents induce magnetic fields. In standard EM this can easily be seen by taking the curl of Amp\`ere's circuital law and substituting the electric field using Faraday's law of induction. One arrives at a wave equation for the magnetic field, $\square \vec B = -\vec \nabla \! \times \! \vec J$, that is sourced by the curl of the current. In curved space and, in addition, in scenarios with non-minimal couplings of EM the situation is more complicated. However, the curl of the standard electric current i.e. the net flux of charged standard model fermions will still act as a source in much the same way.

To solve for the evolution of the electromagnetic fields and the plasma one typically applies Ohm's law. It represents a shortcut to solving Maxwell's equations by providing a link between the current and the electromagnetic fields, $\vec J = \sigma (\vec E + \vec v \! \times \! \vec B)$, where $\vec v$ is the bulk velocity of the plasma and $\sigma$ is the conductivity that has units of an inverse length. Ohm's law derives from the evolution equation of the current and reduces to this simple form after assessing the relevant evolution and interaction timescales. For a general relativistic derivation of the current evolution and a generalised Ohm law for certain situations (see ref.~\cite{Kandus:2007ap} and references therein).

In this section we carefully review the evolution of currents and derive a generalised Ohm law relevant on cosmological scales and also valid in the context of non-minimally coupled EM. We follow the derivation of the current evolution equation given in ref.~\cite{Kandus:2007ap}, and assess the different interaction terms similarly to refs.~\cite{Hollenstein:2007kg, Kobayashi:2007wd, Takahashi:2007ds, Fenu:2010kh}.

\subsection{Relevant particle species and their interactions}  \label{sec:particles}

We aim at deriving the evolution equation of electrical currents on cosmological scales valid from temperatures $T\approx500$ keV (or $z\approx10^9$) corresponding to electron-positron annihilation all the way down to today. Next to the classical electromagnetic fields described by the Faraday tensor ($F$), the constituents of the plasma that we need to follow are the photons ($r$), the free electrons ($e$), the free protons ($p$) and the neutral atoms such as hydrogen and helium ($H$), that start forming due to recombination. In this discussion we can safely ignore the neutrinos and dark matter. However, ultimately we are only interested in the dynamics of electric currents from electrons and protons, and the evolution of the totality of the massive standard model particles, referred to as baryonic matter (label $b$), not to be confused with actual baryons in the particle physics sense ($B$). 

The photons interact through Thomson scattering with the free electrons and protons, $\cC_{rs}^\mu$, where $s=e,\,p$. The electrons and protons interact in addition through Coulomb scattering, $\cC_{ep}^\mu$, a.k.a.\ Rutherford scattering. Finally, the free charges feel an electromagnetic force, $\cC_{sF}^\mu$, while the neutral atoms only interact gravitationally:
\be
\nabla_\alpha T_r^{\mu\alpha} &=& \cC_{r e}^\mu +\cC_{r p}^\mu \,,
\label{eq:nablaTgamma}
\\
\nabla_\alpha T_p^{\mu\alpha} &=& -\cC_{r p}^\mu +\cC_{pe}^\mu +\cC_{pF}^\mu \,,
\\
\nabla_\alpha T_e^{\mu\alpha} &=& -\cC_{r e}^\mu -\cC_{pe}^\mu +\cC_{eF}^\mu \,,
\\
\nabla_\alpha T_H^{\mu\alpha} &=& 0 \,.
\ee
The conservation equation for the baryonic matter reads
\be
\nabla_\alpha T_b^{\mu\alpha}
= \nabla_\alpha \( T_e^{\mu\alpha} +T_p^{\mu\alpha} +T_H^{\mu\alpha} \)
= -\cC_{rp}^\mu -\cC_{re}^\mu +\cC_{pF}^\mu +\cC_{eF}^\mu \,,
\ee
and the Coulomb scattering drops out.

The energy-momentum tensor of the electromagnetic fields derives from its potentially non-minimal action, see next section. In case of the standard Maxwell theory, $\nabla_\alpha T_F^{\mu\alpha} = - F^{\mu\alpha}j_\alpha$, with the 4-current $j^\mu$. In this section, we consider a Lagrangian with a canonically normalised kinetic term, $-(1/4)F^2$, but take into account possible couplings to some scalar fields. Effectively, this means we can write
\be
\nabla_\alpha T_F^{\mu\alpha} = \cK^\mu - F^{\mu\alpha}j_\alpha \,,
\label{eq:divTF}
\ee
where $\cK^\mu$ combines all contributions to the conservation equation due to the interactions with fields beyond the standard model. In addition, we need to allow for a rescaled coupling to the standard model fields, $\alpha \to \alpha=\bar\alpha(1+\delta\alpha)$, where $\delta\alpha$ depends on the coupled fields. The stringent bounds on the fine-structure constant, $\alpha$, translate into the requirement $\delta\alpha\ll 1$ and very little space-time dependence, such that we may neglect derivatives of $\delta\alpha$. The modified $\alpha$ also enters the Thomson and Coulomb collision cross sections. Clearly, the interaction terms must add up to zero, such that $\sum_s \nabla_\alpha T_s^{\mu\alpha}=0$, if we also include all fields coupled to $F^\mn$.

Let us briefly recapitulate how the energy-momentum transfer terms come about in the above conservation equations. In kinetic theory, the photons are described by their phase-space distribution function that follows a Boltzmann equation with collision terms due to the scattering of photons into and out of a given phase-space element. The Boltzmann equation is expanded into a multipole hierarchy, where the multipoles are integrations of the distribution function over momentum, see e.g.\ \cite{Maartens:1998xg}. The first two multipole equations are the energy-momentum conservation equations (\ref{eq:nablaTgamma}). They take the usual hydrodynamical form. The energy density $\rho_r$, the energy flux $q_r^\mu$, and the anisotropic stress $\pi_r^\mn$ of the photons can be interpreted as effective fluid variables that describe the macroscopic, collective properties of the photons, as opposed to the distribution function that describes the microscopic properties of individual photons. The interaction terms, $\cC_{r e}^\mu$ and $\cC_{r p}^\mu$, reflect the energy-momentum transfer due to the collection of local collisions of the photons with the electrons and protons, respectively. In ref.\ \cite{Maartens:1998xg} they are computed up to second order, see also ref.\ \cite{Uzan:1998mc}. They are split relative to the frame $u^\mu$ as $\cC_{r s}^\mu=\cC_{r s} u^\mu + \cC_{r s}^{\la\mu\ra}$ and for the electrons it is found that
\be
\cC_{r e} &=& n_e \sigma_T 
  \[ \frac{4}{3}\rho_r v_e^2 - v_{e\, \alpha} q_r^\alpha
   +\order{3} \] \,,
\\
\cC_{r e}^{\la\mu\ra} &=& n_e \sigma_T 
  \[ \frac{4}{3}\rho_r v_e^\mu - q_r^\mu 
    + v_{e\, \alpha} \pi_r^{\mu\alpha}
    + \order{3} \] \,,
\ee
where $\sigma_T \equiv 8\pi\alpha^2/(3m_e^2)$ is the Thomson cross-section. The relative velocity $v_e^\mu$ stems from the decomposition of the rest frame 4-velocity w.r.t.\ the generic frame: $u^\mu_e \equiv \gamma_e(u^\mu + v_e^\mu)$ with $u_\mu v_e^\mu = 0$ and $\gamma_e \equiv (1-v_e^2)^{-1/2}$, see appendix \ref{app:covariant} for details. For the protons we have
\be
\cC_{r p}^\mu = \beta^2\, \frac{n_p}{n_e}\, \cC_{r e}^\mu
\,, \qquad \beta \equiv \frac{m_e}{m_p} \,.
\ee
Here the factor $\beta^2\approx 3\times 10^{-7}$ makes up for the mass-dependence of the cross-section and strongly suppresses the proton Thomson term as compared to the electron one. Notice that the momentum transfer rate, $\cC_{re}^{\la\mu\ra}$, is first plus second order in cosmological perturbations and also includes perturbations of the number density $n_e$. The energy transfer rate, $\cC_{re}$, starts only at second order in the relative velocity.

The non-relativistic massive particles are well described by a hydrodynamical treatment with vanishing heat flux, viscosity and anisotropic stress. In other words, their Boltzmann hierarchy truncates after the dipole, and the Coulomb collision term is derived and interpreted in the same way as the Thomson terms discussed above. Following refs.~\cite{Uzan:1998mc} and~\cite{Huba:2011} we find for non-relativistic electron-proton collisions
\be
\cC_{pe} &=& n_e n_p \sigma_C v_{Te} m_e \Big[ v_e^2 - \beta v_p^2 
- (1+\beta)v_{e\,\alpha}v_p^\alpha + \order{3} \Big] \,,
\\
\cC_{pe}^{\la\mu\ra} &=& n_e n_p \sigma_C v_{Te} m_e \( 1 +\beta \)
   \Big[ v_e^\mu - v_p^\mu +\order{3} \Big] \,,
\ee
with the Coulomb cross section $\sigma_C \equiv 4\pi\lambda_{ep} \alpha^2 / T_e^2$ and the thermal electron velocity $v_{Te}=(T_e/m_e)^{1/2}$. The Coulomb logarithm is $\lambda_{ep} = 23 + \frac{1}{2} \ln ( T_e^3 / n_e )$ for the temperatures of interest \cite{Huba:2011}. We find $\lambda_{ep} \approx 34$ at early times before it slowly falls off when the baryons start cooling adiabatically, see below. Again, the momentum transfer is first plus second order in perturbations including density perturbations, while the energy transfer only starts at second order in the velocities. 

Finally, the electromagnetic force terms, $\cC_{sF}^\mu$, are not actually due to collisions, but can be consistently included as force terms in the Boltzmann equation \cite{Uzan:1998mc}. The 4-current of standard model charges, here the free electrons and protons, reads
\be
j^\mu = e\( [n_p-n_e]u^\mu + n_p v_p^\mu - n_e v_e^\mu \) \,,
\label{eq:totalcurrent}
\ee
see appendix \ref{app:current} and particularly eq.\ (\ref{eq:partialcurrent}). Even though allowing for non-minimal EM, we take the interaction between standard model charges and the electromagnetic fields to be canonically normalised, as discussed below eq.\ (\ref{eq:divTF}). This means we can write the electromagnetic force terms in their standard form
\be
\cC_{sF} &=& e_s n_s \, v_s^\alpha E_\alpha \,,
\label{eq:Eforce} \\
\cC_{sF}^{\la\mu\ra} &=& e_s n_s \(E^\mu + \veps^{\mu\ab} v_{s\,\alpha} B_\beta \) \,, \label{eq:Lorentzforce}
\ee
not forgetting that $e_s=\pm\, e$, the charge of particles of type $s$, may depend slightly on the fields that interact non-minimally with $F^\mn$. But again, we can neglect derivatives of $e=\sqrt{4\pi\bar\alpha(1+\delta\alpha)}$. The electric and magnetic fields measured in the generic frame $u^\mu$ are projected from the Faraday tensor as follows
\be
E_\mu = F_{\mu\alpha} u^\alpha \, ,
\qquad
B_\mu = \frac{1}{2} \veps_{\mu\ab} F^\ab \,,
\qquad
F_\mn = 2u_{[\mu}E_{\nu]} + \veps_{\mn\alpha}B^\alpha \,.
\ee
The evolution of the electric and magnetic fields is governed by the (potentially modified) Maxwell equations, but we will not use them for the moment.

Before moving on to the current evolution equation, let us summarise the homogeneous and isotropic thermal evolution of the plasma. The formation of neutral hydrogen and helium due to the protons capturing electrons is followed in terms of the ionisation fraction $x_e$. It relates the average number density of the free electrons to the average baryon number density, $\bar n_B$,
\be\label{xedef}
\bar n_e \equiv x_e\, \bar n_B \,, \qquad 
\bar n_B \equiv \bar n_p + \bar n_H = \bar n_e + \bar n_H \, .
\ee
The Universe is charge neutral on average, so we have set $\bar n_e = \bar n_p$.  The evolution of $x_e$ starts off at 1 at early times, and drops rapidly when the recombination processes are falling out of equilibrium, roughly at $T_{rec} \sim 0.3$ eV. When recombination is nearly completed, around $z\sim 1010$, it settles at
\be\label{xefinal}
x_e^f \approx 1.2 \times 10^{-5}\, \frac{\sqrt{\Omega_m}}{\Omega_b h} \, ,
\ee
until reionisation of the plasma due to the radiation from the first stars increases $x_e$ back up to about 0.1, starting roughly from $z\sim 10$. We employ a numerical fit to the evolution of $x_e$ given in eq.\ (3.201) of ref.\ \cite{Mukhanov:2005sc}. The baryon number density, $\bar n_B$, derives from the constant baryon to photon ratio, $\bar n_B/ \bar n_r = 2.6\times 10^{-8}\,\Omega_b h^2$, and the well known result $\bar n_r = 2\zeta(3) a^{-3}$. The average photon energy density is $\bar\rho_r = (\pi^2/15)a^{-4}$. Finally, the non-relativistic species have negligible average pressure and an energy density of $\bar{\rho}_s \simeq m_s \bar{n}_s$.

To compute the Coulomb collision rate we also need to know the evolution of the baryon  temperature (incl.\ electrons). As long as Thomson scattering is efficient the baryons are in good thermal equilibrium with the radiation. Then we have $T_b\approx T_r=T_\text{cmb}/a$ with the CMB temperature today $T_\text{cmb}\approx 2.726\,$K. However, after sufficient dilution of the plasma, the electrons do not meet enough photons anymore to keep thermal contact, and the baryons start to cool adiabatically. When this happens can be seen from the evolution equation for the baryon temperature 
\be\label{Tevol}
\dot T_b + 2H T_b
= \frac{8}{3}\, \frac{\mu}{m_e}\, \frac{\bar{\rho}_r}{\bar{\rho}_b}\, \sigma_T \bar{n}_e
\(T_r - T_b\)\, ,
\ee
with the mean particle mass $\mu$. This evolution equation is derived from the first law of thermodynamics with Thomson scattering as a means of transferring heat from the photons to the baryons, see for instance eq.\ (69) of ref.\ \cite{Ma:1995ey}. With $\bar{\rho}_b\simeq \mu \bar{n}_b = \mu(1+x_e)\bar{n}_B$ and $\bar{n}_e=x_e \bar{n}_B$ we can see that it is the rate
\be
\Gamma_b \equiv \frac{8}{3}\, \frac{\bar{\rho}_r\sigma_T}{m_e}\, \frac{x_e}{1+x_e} \,,
\ee
that governs the thermal evolution of the baryonic matter. It drops below the Hubble rate at a redshift of about $z\sim 120$, when the baryons decouple from the photons and $T_b\propto 1/a^2$.

\subsection{Evolution of currents}  \label{sec:currents}

Let us now write and assess the evolution equation for the spatial part of the total current, $J^\mu\equiv j^{\la\mu\ra}\equiv h^\mu_\alpha j^\alpha$, with the 4-current $j^\mu$ given in eq.~(\ref{eq:totalcurrent}). In appendices \ref{app:velocity} and \ref{app:current} we give details on the derivation. From the energy and momentum conservation equations we derive an evolution equation for the bulk velocity of each species (\ref{eq:vdot1}). These are then used together with the evolution of the number densities (\ref{eq:ndot}) to arrive at the fully general non-linear evolution equation for the total spatial current (\ref{eq:jdotfull}). This equation and its derivation was also given in ref.\ \cite{Kandus:2007ap}, but with a minor confusion about the pressure of the plasma components.

We spell out the equation for the electron-proton system, and insert the electromagnetic force and the elastic Thomson and Coulomb collision terms from above. Even though the  electrons and protons are non-relativistic, we do not a priori neglect their pressure, but rather realise that it only arises at second order in perturbations, $p_s\simeq n_s T_s \ll \rho_s$, for $s=p\,,e$. We replace the pressure perturbations by means of the electron and proton adiabatic sound speeds, $c_e$ and $c_p$, see appendix \ref{app:presspert}. For the energy density, we use $\rho_s = m_s n_s + \order{2}$. Finally, keeping the electromagnetic fields non-perturbative, but neglecting velocity and density perturbations of third and higher order, we arrive at
\be
\dot{J}^{\la\mu\ra} &=& - \frac{4}{3}\Theta J^\mu
 - \Big[\sigma^\mu_{\phantom{\mu}\alpha}+\omega^\mu_{\phantom{\mu}\alpha}\Big]J^\alpha
 - \D_\alpha \big( v_p^\alpha J_p^\mu + v_e^\alpha J_e^\mu \big)
 - e\(n_p-n_e\) \dot{u}^\mu
\nn \\ &&
-e \sigma_C v_{Te} (1+\beta)^2 \Big[ n_p J_e^\mu + n_e J_p^\mu \Big]
- \frac{4 \sigma_T}{3m_e} \Big[ \rho_r\( J_e^\mu +\beta^3 J_p^\mu \)
  + \bar\rho_r J_r^\mu \Big]
\nn \\ &&
+ e^2 \[ \frac{n_e^2}{\rho_e} + \frac{n_p^2}{\rho_p} \]E^\mu
+ \frac{e}{m_e} \Big[v_e^\alpha J_e^\mu -\beta v_p^\alpha J_p^\mu\Big] E_\alpha
- \frac{e}{m_e}\veps^{\mu\ab}\Big[ J_{e\,\alpha} -\beta J_{p\,\alpha}\Big] B_\beta
\nn \\ &&
-c_p^2\[ \big(\Theta+\D_\alpha v_p^\alpha\big) J_p^\mu 
  +\D^\mu\big(v_{p\,\alpha} J_p^\alpha\big) \]
-c_e^2\[ \big(\Theta+\D_\alpha v_e^\alpha\big) J_e^\mu 
  +\D^\mu\big(v_{e\,\alpha} J_e^\alpha\big) \]
\nn \\ &&
-e n_p c_p^2 \frac{\D^\mu \rho_p}{\rho_p} +en_e c_e^2 \frac{\D^\mu \rho_e}{\rho_e}
+ \order{3} \,.
\label{eq:jdot2}
\ee
Here the partial electron and proton currents are $J_e^\mu \equiv -en_ev_e^\mu$ and $J_p^\mu \equiv en_p v_p^\mu$, respectively. The effective Thomson photon current is defined as
\be
J_r^\mu \equiv \frac{3}{4\bar\rho_r} \, \Big[ e(n_e - \beta n_p)q_r^\mu
-\(J_e^\alpha +\beta^3 J_p^\alpha\) \pi_{r\ \alpha}^\mu \Big] \,,
\label{eq:Jr}
\ee
where $q_r^\mu$ and $\pi_r^{\mn}$ are the photon energy flux and anisotropic stress, respectively.

The general relativistic current evolution equation (\ref{eq:jdot2}) includes all effects that arise up to second order in perturbations, also taking number density fluctuations into account. The first line shows all geometric and kinetic effects, comprising the isotropic, shear and vortical expansion, non-linear plasma inhomogeneities, and the acceleration of local charge fluctuations. The second line describes the effect of Coulomb and Thomson scattering on the total current. The third line shows how electric fields generate currents, and it accounts for the Hall effect, a coupling between the current and the magnetic field. Finally, the last two lines in eq.\ (\ref{eq:jdot2}) are due to the pressure perturbations (and their transformation into the generic frame $u^\mu$ that we employ to describe the current and the electromagnetic fields). In most treatments of magnetic field generation in the recombination era the pressure perturbations have been neglected \cite{Matarrese:2004kq, Kobayashi:2007wd, Fenu:2010kh}, even though they are known to generate magnetic fields on small scales, an effect known as Biermann battery, see e.g.\ \cite{Widrow:2011hs}.

To better understand the evolution of the current, let us rewrite and simplify eq.\ (\ref{eq:jdot2}) a bit. For this purpose we define the average number density $\bar n=\bar n_e=\bar n_p=x_e \bar n_B$, and introduce the Coulomb rate, $\Gamma_C$, the Thomson rate, $\Gamma_T$, and the plasma frequency, $\omega_p$,
\be
\Gamma_C  \equiv  e \bar n \sigma_C v_{Te}  \,,\qquad
\Gamma_T  \equiv  \frac{4\bar \rho_r \sigma_T}{3m_e}  \,,\qquad
\omega_p  \equiv  \( e^2 \bar n / m_e \)^{1/2}  \,.
\ee
We change variables for the partial currents and number densities by defining the centre of mass velocity and number density fluctuation
\be
v_{pe}^\mu \equiv \frac{ v_p^\mu +\beta v_e^\mu }{1+\beta}
\,,\qquad
\delta_{pe} \equiv \frac{ n_p +\beta n_e }{\bar n(1+\beta)} -1 \,.
\ee
It is often assumed that the centre of mass velocity and density coincide with the velocity and density of the baryonic matter. However, we point out that this is not necessarily correct at second order and needs to be checked (numerically) by treating the electrons, protons, and neutral atoms separately to compare $v_{pe}^\mu$ with $v_H^\mu$ and $v_b^\mu$, etc. Next we define the centre of mass current and the local charge number fluctuation
\be
J_{pe}^\mu \equiv e \bar n\(1+\delta_{pe}\) v_{pe}^\mu
\,, \qquad
\Delta \equiv \frac{n_p - n_e}{\bar n} \,.
\ee
Using these definitions, we replace the velocities, the partial currents and number densities in eq.\ (\ref{eq:jdot2}) and keep terms up to second order in perturbations. The terms that come from the pressure perturbations are strongly suppressed even though $c_s^2\Theta J_s^\mu$ and $c_s^2\D^\mu\rho_s$ come in linearly. They are effectively second order contributions because the sound speed is so small, $c_e^2 \equiv \dot{\bar\rho}_e / \dot{\bar p} \simeq (5/3)\,T_b / m_e \approx 10^{-5} (10^{-4}/a)$ and $c_p^2 = \beta c_e^2$ from eq.\ (\ref{Tevol}). Therefore, we can safely neglect the terms $c_s^2[\, (\D_\alpha v_s^\alpha)J_s^\mu +\D^\mu(v_{s\,\alpha} J_s^\alpha)\,]$ as compared to the similar terms in the first line of eq.\ (\ref{eq:jdot2}). The remaining pressure gradients act as sources for the current but are only relevant deep in the  non-linear regime. Next, to facilitate the qualitative discussion we also consider the electric and magnetic fields as being small fluctuations. We neglect terms $\order{v^2 E,\, \Delta^2 E,\, \Delta v B}$, even though these may have to be kept for a fully consistent treatment \cite{Mongwane:2012gg}. Finally, we neglect $\beta = m_e/m_p$ as compared to 1. With these substitutions and simplifications we arrive at
\be
\dot{J}^{\la\mu\ra} &=& - \frac{1}{3}\(4+3c_e^2\)\Theta J^\mu
 - \Big[ \sigma^\mu_{\phantom{\mu}\alpha} 
       + \omega^\mu_{\phantom{\mu}\alpha} \Big] J^\alpha
 -\frac{1}{e\bar n}\D_\alpha\Big(2J_{pe}^{(\alpha}J^{\mu)}-J^\alpha J^\mu\Big)
 - e\bar n \Delta \dot{u}^\mu
\nn \\ &&
- \Gamma_C \Big[ (1+\delta_{pe}) J^\mu - \Delta J_{pe}^\mu \Big]
- \Gamma_T \Big[ (1+\delta_r-\beta\delta_{pe}) J^\mu 
               - (1+\delta_r) J_{pe}^\mu + J_r^\mu \Big]
\nn \\ &&
+ \omega_p^2 \Big[ 1 + \delta_{pe} - \Delta \Big] E^\mu
- \frac{\omega_p^2}{e\bar n} \veps^{\mu\ab} \Big[ J_\alpha 
  - J_{pe\,\alpha}\Big] B_\beta
\nn \\ &&
+ c_e^2 \Theta J_{pe}^\mu
+ e \bar n c_e^2 \D^\mu \Big( \delta_{pe} - \Delta \Big)
+ \order{3} \,.
\label{eq:jdot2_simp}
\ee
where $\delta_r \equiv \rho_r/\bar\rho_r -1$ is the photon density contrast. In this approximation the effective photon current reads
\be
J_r^\mu \equiv \frac{3}{4\bar\rho_r} \, \Big[ 
e\bar n(1+\delta_{pe}-\Delta)q_r^\mu
+\( J_{pe}^\alpha - J^\alpha \) \pi_{r\,\alpha}^\mu \Big] \,,
\ee
In eq.\ (\ref{eq:jdot2_simp}) we can see that both Coulomb and Thomson scattering damp currents as well as drive them through the terms $\propto J_{pe}^\mu$ and $J_r^\mu$. Pressure gradients and electric fields generate currents and magnetic fields curl them. Finally, the pressure gradients, here the electron sound speed $c_e^2$ times density gradients, act as sources for the current. However, once one takes the curl to compute the source in $\square \vec B = -\vec \nabla \! \times \! \vec J$, it is clear that the pressure is only relevant in the non-linear regime where there is vorticity, as $(\text{curl}\,\D f)_\mu = -2\dot f \omega_\mu$ \cite{Kobayashi:2007wd}. Known as the Biermann battery, their effect during recombination remains to be quantified as it was neglected in ref.\ \cite{Fenu:2010kh} and the other relevant literature.

We remind the reader that the only impact of non-minimal couplings of EM fields on the evolution of standard model currents is an effective space-time dependence of the fine-structure constant and therefore the electric charge. We did not use the Maxwell field equations, which is the reason why non-standard currents do not appear in the current evolution equation. In fact, we did not use the Einstein field equations either, and therefore these results also hold in scenarios with modifications to gravity, such as the presence of couplings of EM to curvature invariants.

\subsection{Generalised Ohm's law}  \label{sec:ohmfinal}

\begin{figure}[tb]
\centering
\includegraphics[width=0.7\textwidth]{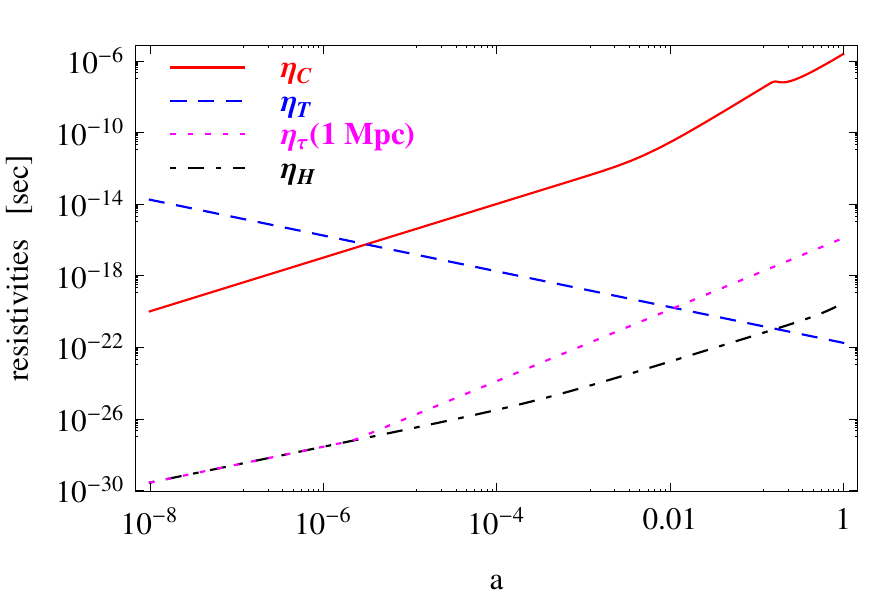}\\
\includegraphics[width=0.7\textwidth]{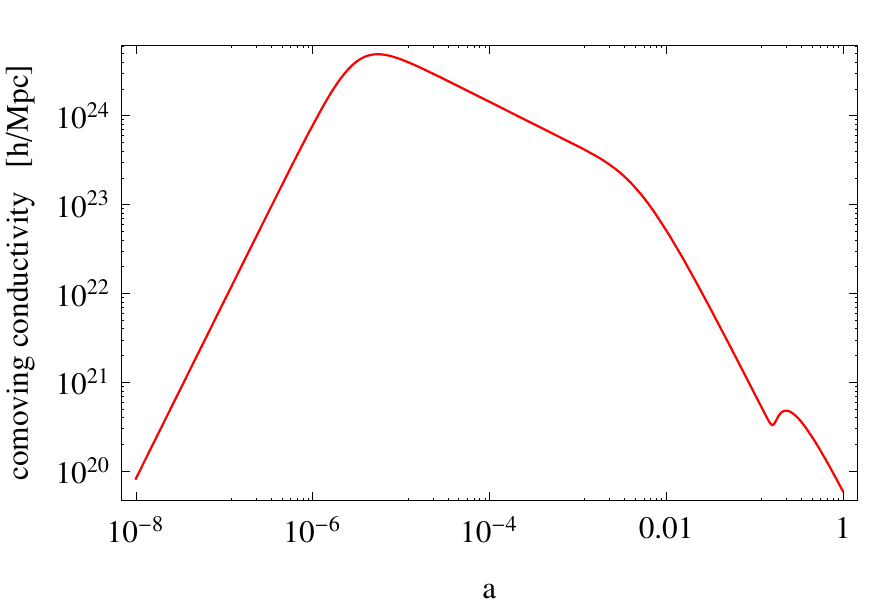}
\caption{\emph{Upper panel:} we compare the various resistivities, defined in eq.~(\ref{eq:resist}) for a comoving correlation length of $1\, \Mpc$, that are relevant in the cosmological Ohm law~(\ref{eq:genOhm}). Notice that $\eta_T\gg\eta_C$ for $a\lesssim 3\times 10^{-6}$, well before recombination, and then $\eta_T\ll\eta_C$ until today.
\emph{Lower panel:} we plot the comoving conductivity $\hat\sigma\equiv a/(\eta_C+\eta_T)$.}
\label{fig:resist}
\end{figure}

Let us now estimate the different timescales involved in the evolution of currents. The main features of eq.\ (\ref{eq:jdot2_simp}) remain if we linearise it in a FLRW background. The proton-electron centre of mass current, $J_{pe}^\mu$, is only present because we worked in an arbitrary frame, $u^\mu$. Let us now choose $u^\mu$ to be the centre of mass frame, which is very close to the baryon frame, then $J_{pe}^\mu\equiv 0$. We express the spatially projected 4-vectors with respect to a comoving basis, i.e.\ $J^\mu=a^{-1}\(0,\,\VJ\)$ etc. Then we find the following linear current evolution equation
\be
\dot{\VJ} + \( 4H + \Gamma_C + \Gamma_T\) \VJ = - \Gamma_T\VJ_r + \omega_p^2\VE\,,
\ee
where the Hubble parameter $H$ comes from $\Theta=3H+\order{1}$. To write a generalised cosmological Ohm law in a familiar form, we approximate the time derivative of the current by a characteristic timescale of the problem, $\tau$, through $\dot \VJ \simeq \tau^{-1} \VJ$. For large-scale fluctuations of physical correlation length $L$ larger than the Silk damping scale, typically $\tau \sim \min(L,\, H^{-1})$. Then we can write the linear cosmological Ohm's law as
\be
\(\eta_\tau + 4\eta_H + \eta_C + \eta_T\) \VJ + \eta_T \VJ_r \simeq \VE \,,
\label{eq:genOhm}
\ee
where the various resistivities are 
\be
\eta_\tau \equiv \frac{\tau^{-1}}{\omega_p^2} \,, \qquad
\eta_H \equiv \frac{H}{\omega_p^2} \,, \qquad
\eta_C \equiv \frac{\Gamma_C}{\omega_p^2} \,, \qquad
\eta_T \equiv \frac{\Gamma_T}{\omega_p^2} \,,
\label{eq:resist}
\ee
and have the dimensions of time. The resistivities quantify the efficiency of generating currents by the presence of an electric field. We compare the resistivities as functions of the scale factor in the upper panel of figure \ref{fig:resist} and notice that $\eta_T\gg\eta_C$ for $a\lesssim 3\times 10^{-6}$, well before recombination, and then $\eta_T\ll\eta_C$ until today. The resistivity due to the expansion, $\eta_H$, is small, being proportional to the Hubble parameter and overcome by the much faster interaction rates. The characteristic evolution timescale, $\eta_\tau$, is likewise negligible since it is inversely proportional to the large scales we consider, i.e.\ the time derivative in the current evolution equation can safely be neglected. We can write eq.~(\ref{eq:genOhm}) in the form
\be
\VJ + \sigma_E \eta_T \VJ_r \simeq  \sigma_E \VE \,,
\label{eq:genOhmconduct}
\ee
where $\sigma_E \equiv \( \eta_C + \eta_T \)^{-1}$ is the electric conductivity of the plasma with the dimensions of an inverse length. 

On the other hand, we should regard the effective photon current, $\VJ_r$, as the external source, rather than the electric field. We need to compare $\VJ$ and $\sigma_E\VE$ to see which one is generated by the photon current. To do so we have to use the Maxwell equations. Let us assume standard EM for the rest of this discussion. Because of conformal invariance we can rescale the electromagnetic fields and the currents according to
\be
\VEtilde \equiv a^2 \VE  \,,\qquad
\VBtilde \equiv a^2 \VB  \,,\qquad
\VJtilde \equiv a^3 \VJ  \,,\qquad
\VJtilde_r \equiv a^3 \VJ_r  \,,
\ee
and use the flat space-time Maxwell equations for the twiddled vectors: $\nabla\!\cdot\VBtilde=0$, $\nabla\!\cdot\VEtilde=a^3Q$, $\nabla\!\times\VBtilde=\VEtilde' + \VJtilde$, and $\nabla\!\times\VEtilde=-\VBtilde'$. The prime denotes conformal time derivative. From the curl of Amp\`ere's law together with the induction equation we find the sourced wave equation $\VBtilde''-\nabla^2\VBtilde = \nabla\!\times\!\VJtilde$. Finally, we take the curl of eq.~(\ref{eq:genOhmconduct}) and use the sourced wave equation to find
\be
\VBtilde''-\nabla^2\VBtilde +\hat\sigma \VBtilde'
\simeq - \frac{\hat\sigma \eta_T}{a} \nabla\times\VJtilde_r \,,
\label{eq:bwave}
\ee
where the comoving conductivity is $\hat\sigma\equiv a\sigma_E$, see figure \ref{fig:resist}. On comoving scales larger than ${\hat\sigma}^{-1}$ one concludes that the first two terms on the left hand side can be neglected w.r.t.\ the third term. This is equivalent to dropping $\VJ$ with respect to $\sigma_E\VE$ in eq.\ (\ref{eq:genOhmconduct}). In this case eq.\ (\ref{eq:bwave}) can be integrated to compute the magnetic field, which was done in ref.\ \cite{Fenu:2010kh}. They find the photon current to be an active source through recombination until about $z\sim 300$, even though Thomson scattering starts to become inefficient already well before recombination, see figure \ref{fig:resist}. Magnetic fields of $\approx 10^{-24}\,\Gauss$ on large scales are found, see the beginning of next section.

Finally, one may ask whether the cosmological Ohm's law is applicable on very large and super-horizon scales. The resistivities that relate the current to the electric field are due to scattering processes. Thus, one might conclude that the resistivities are not well defined on super-horizon scales due to causality -- particles in different Hubble volumes are never in causal contact with each other. However, let us emphasise that the general current evolution equation (\ref{eq:jdot2}) was derived from the first two multipole equations of the Boltzmann hierarchies for the photons, electrons and protons. The collision and EM force terms result from passing from the microscopic kinetic theory to the macroscopic fluid description of the plasma, through integration over momentum space. They are local, space-time dependent terms that describe the local energy-momentum transfer rates for the effective fluid description, and not interactions between spatially separated particles. Assume that inflation has imprinted super-horizon correlations in the primordial curvature perturbations which further induce non-trivial correlations in the photon, electron and proton distributions but did not generate any magnetic fields. Around recombination, the local momentum transfer between the photons, electrons and protons leads to generation of currents and therefore EM fields. Even though they are generated causally, they inherit the super-horizon correlations from their sources. Despite, causality being built into the relativistic Boltzmann hierarchy, the initial conditions are generically acausal. In other words, inflation ensures that the charge flux in two disconnected Hubble regions is correlated intrinsically. Therefore, the EM fields generated by the correlated currents will be correlated on super-Hubble scales as well. Consequently, the resistivities, the conductivity and Ohm's law are still well defined on super-horizon scales. 

Notice that in the case of non-minimal EM the form of the generalised Ohm law would not change, as we have not used the Maxwell field equations to derive it. Only, the fine-structure constant that goes into the resistivities would become slightly space-time dependent, as we considered a canonically normalised theory.


\section{Scenarios of modified electromagnetism}\label{sec:modified}


The original motivation for looking at the possibility of late-time couplings between EM and cosmological (pseudo-)scalars came from explaining the observed cosmological magnetisation. Some time after the completion of recombination the Universe is populated by seed fields whose stochastic distribution is characterised by a power spectrum of the form~\cite{Fenu:2010kh}
\be\label{sshape}
  k^3P_S \propto \left\{ \begin{array}{ll}
    k^8 & \quad \text{for}\ k\ll k_{eq} \\
    k   & \quad \text{for}\ k\gg k_{eq} \end{array}\right. \, ,
\ee
where $k_{eq}=\cH_{eq}\approx 0.01\,h/\Mpc$ is the mode that enters the horizon at matter-radiation equality. This spectral shape is reached roughly at redshift $z\sim 300$, when the non-linear processes saturate and the magnetic fields continue their evolution through adiabatic expansion, $B\propto a^{-2}$, in the absence of any other couplings and effects.  At $a=10a_{eq}$, where $a_{eq}\equiv\Omega_r/\Omega_m\approx 1/3240$, the physical field strength on a scale $k=20k_{eq}$ is found to be $\approx 10^{-24}\,\Gauss$, see right panel of figure 4 in ref.\ \cite{Fenu:2010kh}.  Clearly, these fields are not going to be strong enough if they are to explain observations. One can then think of adding cosmological (pseudo-)scalar fields to this picture, whose interactions with EM might act only at late times and perhaps boost these weak seeds to more reasonable amplitudes.  What we will find is that, even in the emptiest Universe, the very low resistivity of the plasma will prevent meaningful modifications of Maxwell's theory from producing any effect.

We thus consider a scalar or pseudo-scalar field $\phi$ (be it the quintessence, dilaton, axion or any other effective cosmological field) with generic Lagrangian density $\cL_\phi\equiv\cP(X,\phi)$ and kinetic term $X\equiv-\frac{1}{2}\partial^\alpha\phi\partial_\alpha\phi$ that is minimally coupled to gravity but non-minimally to EM, as given in the following general action~\cite{Grasso:2000wj,Giovannini:2003yn,Subramanian:2009fu}
\bea\label{intL}
  S[\phi, A_{\mu}] \equiv \int\dd^4x\, \sqrt{-g} && \[
  \frac{\mpl^2}{2}R +\cP(X,\phi)
  -\frac{1}{4}f^2(\phi)F_\ab F^\ab
  \right. \nonumber \\ &&\left.
  \ \ +\, \frac{1}{4}g(\phi)F_\ab \tilde{F}^\ab
  -\frac12 m^2(\phi)A_\alpha A^\alpha + j_\alpha A^\alpha \] \, .
\eea
Here $F_{\mu\nu} \equiv \partial_\mu A_\nu-\partial_\nu A_\mu$ is the EM field tensor and $\tilde{F}^{\mu\nu}$ is its dual, defined as $\tilde{F}^{\mu\nu}\equiv\frac{1}{2}\eta^{\mu\nu\alpha\beta}F_{\alpha\beta}$ with $\eta^{\mu\nu\alpha\beta}$ being the totally antisymmetric Levi-Civita tensor with $\eta^{0123}=+|g|^{-1/2}$.  The coupling functions $f(\phi)$ and $g(\phi)$ are dimensionless while $m(\phi)$ represents an effective mass for the photon.

Since the canonical normalisation of the kinetic term of the vector potential leads to a redefinition of the electron charge as $e\rar e/f$, the term $f^2(\phi)$ is directly responsible for the spatiotemporal variation of the fine-structure constant $\alpha$.  The axial coupling function $g(\phi)$ arises naturally in theories where a pseudo-scalar is present, such as the axion. It is modelled on the analogue anomalous coupling between photons and the axial neutral mesons of QCD.  Note that we only consider the field coupling to EM and not the derivative couplings as they will presumably be suppressed by the scale associated with the derivative interaction.

A late-time coupling promoted by the Lagrangian~(\ref{intL}) is rich of phenomenological consequences.  A late-time running of $\alpha$ was studied in connection with DE in a series of works, the latest being~\cite{Thompson:2012pj} (see also~\cite{Avelino:2006gc} for an extensive list of references).  In the comprehensive review~\cite{Uzan:2010pm} a detailed account of such effects, and including early Universe ones, can be found.  An axial coupling to $F\tilde{F}$ gives rise primarily to polarisation rotation and distortion effects, which crucially depend on the nature of the (pseudo-)scalar field.  This applies equally well to light coming from the CMB as well as that from very distant and bright objects such as quasars.  Works which can guide through the literature in the subject are~\cite{Ni:2011ti,Urban:2010wa,Payez:2008pm}.  Finally, we will see that the limits on the mass of the photon are extremely severe which means that any observational signatures, such as CMB distortions, are consequently negligible.

The Euler-Lagrange equations in compact form are
\be\label{EOMfull}
f^2 \partial^\alpha F_{\alpha\mu} + 2f \partial^\alpha f F_{\alpha\mu} + \partial^\alpha g \tilde F_{\alpha\mu} - m^2 A_\mu + j_\mu = 0 \, .
\ee
For the current, we assume a simple Ohm law, $j_\mu = \sigma_E E_\mu = \sigma_E (0,-A_i'/a)$, and define the comoving conductivity again by $\hat\sigma\equiv a\sigma_E$. Here we neglect the effective photon current discussed in section \ref{sec:ohmfinal}, as it is irrelevant for $z\lesssim 300$. The scalar or pseudo-scalar field $\phi$ is assumed to be a spatially homogeneous and isotropic background for the EM field. In Coulomb gauge, $\partial^i A_i = 0 = A_0$, the Euler-Lagrange equations then become
\be\label{EOMexp}
f^2 A_i'' - f^2 \nabla^2 A_i + 2 f f' A_i' + g' \epsilon_{ijk} \partial_j A_k + a^2 m^2 A_i + \hat \sigma A_i' = 0 \, ,
\ee
where $\epsilon_{123} = + 1$. Notice that in the case of a non-trivial $f(\phi)$ the value of the conductivity can or can not  have an explicit factor of $f^2$ as well, depending on whether the factor $f^2$ embraces the entire Lagrangian including the photon vertex or only the kinetic term $F^2$ as in eq.\ (\ref{intL}).

Before analysing the impact of the modifications to EM at late times, it is instructive to look at the standard mode equation for $f(\phi)=1$, $g(\phi)=0=m(\phi)$. Upon transforming to Fourier space, the vector potential is expressed in the helicity basis, $\Veps^\Vk_h$ with $h=\pm$ and we can readily write down the mode equation as
\be\label{modesZero}
\cA_{h}'' + \hat\sigma \cA_{h}' + k^2 \cA_{h} = 0 \, .
\ee
Assuming the friction term to be time-independent for simplicity (the reasons why this assumption does not affect the general results will become clear through the section), we can write the solution as
\be\label{cAzeroth}
\cA_{h} \approx C_1\, e^{-\frac12 \left\{ 1+\sqrt{1-4\kappa^2} \right\} \hat\sigma\eta} + C_2\, e^{-\frac12 \left\{ 1-\sqrt{1-4\kappa^2} \right\} \hat\sigma\eta} \, ,
\ee
where $\kappa \equiv k/\hat\sigma$. We can study the behaviour of the solution in the two separate regimes, $\kappa \ll 1$ and $\kappa \gg 1$, which leads to 
\be
& \cA_{h} \approx C_1\, e^{-\hat\sigma\eta} + C_2\, e^{-\kappa^2\hat\sigma\eta} \qquad & \text{for} \quad 2 \kappa \ll 1\label{cAsmallk} \,,
\\
& \cA_{h} \approx e^{-\hat\sigma\eta/2} \left(C_1\, e^{-ik\eta} + C_2\, e^{ik\eta}\right) \qquad & \text{for} \quad 2 \kappa \gg 1 \label{cAlargek} \,.
\ee
We recover the known result that the vector potential is either essentially frozen in the plasma or a rapidly oscillating and decaying wave, respectively. The transition between these two regimes is given by the comoving conductivity of the plasma $\hat\sigma$.  If we are interested in the Universe around and after recombination until today, it is easy to see that the conductivity is tremendously large, being still $\hat\sigma \approx 10^{20}\, h/\Mpc$ at $z=1$, see lower panel of figure~\ref{fig:resist}.


\subsection{Running fine-structure constant}

We focus here on the case with $g(\phi) = m(\phi) = 0$ and study the effects of the term $f^2(\phi)F_\ab F^\ab$ in the action. In the context of inflation, this model was studied for example in refs.\ \cite{Martin:2007ue, Demozzi:2009fu}. The Fourier space equation we want to solve is thus
\be
f^2 \cA_{h}'' + \hat\sigma \cA_{h}' + 2 f f' \cA_{h}' + f^2 k^2 \cA_{h} = 0 \,,\label{eomRal1} \\
\hat\cA_{h}'' + \hat\sigma \hat\cA_{h}' - \frac{f''}{f} \hat\cA_{h} - \hat\sigma \frac{f'}{f} \hat\cA_{h} = 0 \label{eomRal2} \, ,
\ee
where in the second line we have defined $\hat\cA_{h} \equiv f\cA_{h}$ and have also made the substitution $\hat\sigma\rar f^2\hat\sigma$ to exemplify the scenario where the factor $f^2$ appears in front of the entire Lagrangian.

In the latter case we can imagine solving this equation for $f'/f$ nearly constant (and negligible $f''/f$) so that it reduces to something very similar to the previous case.  This is realised by choosing $f = f_0 \exp\{\lambda \hat\sigma(\eta-\eta_0)\}$ where $\lambda$ is a dimensionless constant, $\eta_0$ is conformal time today, and we chose $f_0 = f(\phi(\eta_0)) = 1$.  The solutions to eq.\ (\ref{eomRal2}) now read
\be\label{cAfirst}
\cA_{h} = C_1\, e^{-\frac12 \left\{ 1+2\lambda+\sqrt{(1+2\lambda)^2-4\kappa^2} \right\} \hat\sigma\eta} + C_2\, e^{-\frac12 \left\{ 1+2\lambda-\sqrt{(1+2\lambda)^2-4\kappa^2} \right\} \hat\sigma\eta} \, .
\ee
We see immediately that on large scales $k\ll1/\Mpc$ the solutions are going to behave in much the same way as in the case without coupling, as long as $\lambda > -1/2$.
Furthermore, if we imagine to retain only the coupling $f^2 F^2$ and not the $f^2 j_\alpha A^\alpha$ then the fine-structure constant will be running with the scalar field. For a similar $f$, the running is given by
\be
\frac{\Delta\alpha}{\alpha} = -2\frac{\Delta f}{f} \approx 2 \left|\lambda\hat\sigma\Delta\eta \right| \,.
\ee
An overview on the upper bounds on the variation of $\alpha$ can be found in ref.\ \cite{Uzan:2010pm}. In any case, even a conservative limit such as $\Delta\alpha/\alpha \leq 10^{-5}$ poses a very strong constraint on the effects of the coupling under inspection, i.e.\ $|\lambda| \lesssim 10^{-25}\,(\Delta\eta/\Mpc)^{-1}$.


\subsection{Running axial coupling}

This case corresponds to $f(\phi) = 1$ and $m(\phi) = 0$. During inflation, this was investigated for example in refs.\ \cite{Anber:2006xt, Durrer:2010mq, Barnaby:2010vf, Byrnes:2011aa}. The equation of motion for the vector potential now reads
\be\label{eomaxial}
\cA_{h}'' + \hat\sigma \cA_{h}' + \left( k^2 + h k g' \right) \cA_{h} = 0 \, .
\ee
Obviously, one can study this equation in great detail by numerical means. But as in the previous case, for an order of magnitude estimate it is sufficient to look for analytical solutions for roughly constant $g_0\equiv g'$. We define $\xi \equiv g_0 / \hat\sigma$ and find
\be\label{cAsecond}
\cA_{h} = C_1\, e^{-\frac12 \left\{ 1+\sqrt{1-4h\kappa\xi-4\kappa^2} \right\} \hat\sigma\eta} + C_2\, e^{-\frac12 \left\{ 1-\sqrt{1-4h\kappa\xi-4\kappa^2} \right\} \hat\sigma\eta} \, .
\ee

An axial coupling of this kind gives rise to cosmic birefringence and dichroism, see for instance ref.\ \cite{Ni:2007ar}. Hence, the most important effect is the coherent rotation of the linear polarisation vector for polarised light that is travelling to us from the last scattering surface or from distant objects such as quasars~\cite{Alighieri:2010pu}.  Current observations limit such effects to be of the order of about a degree for CMB photons~\cite{Gruppuso:2011ci} or light from distant quasars~\cite{Alighieri:2010eu}.  In terms of the coupling $g$, the change in the polarisation rotation angle $\theta$ is given by
\be\label{cbest}
\Delta \theta \propto \Delta g(\phi(\eta)) = g(\eta_\text{source}) - g(\eta_0) \simeq g' \Delta\eta \,.
\ee
For instance, taking the CMB limit alone, the allowed average rotation is $\Delta\theta \lesssim 0.02\,$rad.  With $\Delta\eta \approx 14000\,\Mpc$ from the last scattering surface to today, we realise that $g_0 \lesssim 10^{-6}\,\Mpc^{-1}$. For an average conductivity of $\hat\sigma\sim 10^{22}\,\Mpc^{-1}$, we have $\xi \lesssim 10^{-28}$ in eq.\ (\ref{cAsecond}) and any effect of this coupling evanesce due to the much more important conductivity friction term.

In principle, one can imagine to get around this problem by employing some kind of resonance in the coupling~\cite{Byrnes:2011aa}.  As a representative example, let us consider the case of an oscillating coupling of the form $g'(\eta) = g_0 \sin (\omega \eta)$, where $\omega$ represents the oscillation frequency (in conformal time). To grasp the broad features of such a choice, we can roughly approximate the sine with a square wave and use the solutions for constant $g'$ as before.  The solutions now grow periodically for one half of the cycle and decay for the other half with alternating cycles for opposite helicities.  If, for example, there is a slight inequality of ups and downs one of the helicity modes will be amplified as compared to the other, on average. A half-cycle lasts for $\eta_c\sim \pi / \omega$, and from eq. (\ref{cAsecond}) we see that the net amplification within a half-cycle is $\exp\{ -h \kappa \xi \hat\sigma \eta_c\}$. For a source that is $\Delta\eta$ away from the observer, a photon goes through about $n_c = \Delta\eta / \eta_c$ half-cycles, and the typical difference between ups and downs is $1/\sqrt{n_c}$. All together we obtain an effective exponential growth given by
\be\label{solOscGrow}
|\cA_{h}| \approx C_2 e^{-h \pi^{3/2} \kappa g_0 / (\omega\sqrt{\omega \Delta\eta}) } \, .
\ee
Focusing on large wavelengths, for instance 1 Mpc, then $k/\hat \sigma \sim 10^{-22}$, while for CMB photons we still have $g_0\lesssim 10^{-6}\,\Mpc^{-1}$. To make the exponent larger than unity for the negative helicity mode, we would need a tiny $\omega$:
\be
\omega < \frac{\pi}{\Delta\eta} \(\frac{k}{\hat\sigma} g_0 \Delta\eta \)^{2/3}
\sim 10^{-20}\,\Mpc^{-1} \,.
\ee
The conclusion is that there is essentially no oscillation and we are back to constant $g'$.


\subsection{Photon mass}

In this case, with the choice $f(\phi) = 1$ and $g(\phi) = 0$, we allow for a non-zero photon mass $m$. For a treatment of massive photons during inflation, see for instance ref.\ \cite{Demozzi:2009fu}. Defining $\hat m \equiv a m$, the equations of motion read
\be
\cA_{h}'' + \hat\sigma\cA_{h}' + \left(k^2 + \hat m^2\right) \cA_{h} &=& 0
\label{eommass} \\
\chi'' + \hat\sigma\chi' + 2k^2 \frac{\hat m'}{\hat m^3} \chi' + \left(k^2 + \hat m^2\right) \chi &=& 0
\label{eomchi} \, .
\ee
The second equation governs the evolution of the longitudinal degree of freedom $\chi$, which necessarily appears when the photon acquires a mass. In the simplest case of a constant mass term the solutions for the three polarisations coincide
\be\label{cAthird}
(\cA_{h},\, \chi) = C_1\, e^{-\frac12 \left\{ 1+\sqrt{1-4\kappa^2-4\mu^2} \right\} \hat\sigma\eta} + C_2\, e^{-\frac12 \left\{ 1+\sqrt{1-4\kappa^2-4\mu^2} \right\} \hat\sigma\eta} \, ,
\ee
where $\mu \equiv \hat m / \hat \sigma = m/\sigma_E$.

A recent review for constraints on the photon mass is ref.~\cite{Goldhaber:2008xy}, where from we learn that a conservative limit on $\hat m$ today (which is $m$ itself) is $\hat m < 10^{-18}$ eV which corresponds to $10^{11}\,\Mpc^{-1}$. This means that $\mu \lesssim 10^{-9}$ today. Once again, we conclude that the conductivity is too large for the photon mass to have any sizeable effect on the solutions. Things would hardly be different for a slow time-variation of $\hat m$, for the only term where this intervenes in eq.\ (\ref{eomchi}) is suppressed by a factor of $(k/\hat m)^2$.


\section{Summary and conclusions}\label{TheEnd}


Large-scale magnetic fields are present not only in bound cosmic structures such as galaxies and clusters, but also they appear to fill a large fraction of the intergalactic space. This fact is difficult to explain by means of standard astrophysical processes. Also the early Universe mechanisms of magnetogenesis are mostly not viable or not efficient, even if one allows for non-minimal EM. On the other hand, the non-linear dynamics of the cosmic plasma in the recombination era do source large-scale magnetic fields, even though with insufficient amplitudes. Taking this mechanism as a promising starting point, we have carefully revisited the non-linear dynamics in a more general setting, allowing for non-minimal couplings of EM to (effective) scalar fields, in the hope of boosting the magnetic seeds.

We have derived the fully general and non-linear evolution equation of the current, and specialised it to the photon-electron-proton plasma, relevant in the recombination era, taking into account all possible effects up to second order in perturbations. We paid particular attention to allowing for possible effects of non-minimal couplings of EM, refraining from applying the Maxwell and Einstein field equations that we want to modify. Subsequently, we have discussed the relevance of the different effects on the current and derived a cosmological Ohm's law that is valid even in non-minimal EM. We also point out that Ohm's law is valid, and the resistivity and conductivity are well defined, on large cosmological scales, even beyond the Hubble horizon, as long as super-horizon correlations were initially imprinted into the cosmic plasma (through inflation).

We have investigated several different types of couplings of EM to scalar and pseudo-scalar fields. Candidate fields which could do the job are for example the DE field, or a cosmological axion background. We have explored the possibilities of a running fine-structure constant, $f^2(\phi)F^2$, a running axial coupling, $g(\phi)F\tilde{F}$, and an effective photon mass $m^2(\phi)A^2$. In all three cases we conclude that current observational and experimental limits do not allow for additional amplification of EM fields on cosmological scales. Even at scales well beyond a Mpc the conductivity of the plasma is so high that it inhibits any departure from the flux freezing behaviour, resulting in the strength of the magnetic field to be about $3\times10^{-29}\,\Gauss$ today.

\acknowledgments

We would like to thank Chiara Caprini, Alexander Dolgov, Ruth Durrer, Elisa Fenu, Roy Maartens and Kandaswamy Subramanian for helpful discussions. L.H.\ and R.K.J.\ acknowledge financial support from the Swiss National Science Foundation. F.U.\ thanks the 
D\'ep\-artement de Physique Th\'eorique of the Universit\'e de Gen\`eve for its hospitality.

\appendix

\section{Conventions}\label{app:conv}

We work with a metric signature \mbox{($-$ + + +)}. For tensor components, Greek indices take values $0\ldots3$, while Latin indices run from $1$ to $3$.  We employ natural Heaviside-Lorentz units such that $c=\hbar=k_B=\epsilon_0=\mu_0\equiv 1$ and the electric charge is $e=\sqrt{4\pi\alpha}$. The reduced Planck mass is defined as $\mpl=(8\pi G)^{-1/2}$. Where the metric is not specified we apply the 1+3 covariant Ehlers-Ellis formalism \cite{Ellis:1998ct, Tsagas:2007yx}, defining a splitting of space-time and all tensors w.r.t.\ a time-like 4-vector field $u^\mu$, with $u^2=u^\alpha u_\alpha=-1$, referred to as a generic observer frame. If not mentioned explicitly, we do not specify $u^\mu$ any further. The covariant derivative is split into a time derivative parallel to $u^\mu$ denoted by a dot, and a spatial derivative, $\D_\mu$, projected with $h_\mn\equiv g_\mn+u_\mu u_\nu$ orthogonal to $u^\mu$. We mostly follow  ref.~\cite{Clarkson:2010uz} for the notation, which we summarise in appendix \ref{app:covariant}.

When using a Friedmann-Lema\^{\i}tre-Robertson-Walker (FLRW) metric, the Hubble parameter is $H \equiv \dot a/a$ where the over-dot denotes the derivative w.r.t the cosmic time $t$. The conformal Hubble parameter is ${\cal H}\equiv a'/a$, where $'$ denotes the derivative w.r.t the conformal time $\eta$ defined as $dt=a d\eta$, $a$ being the scale factor normalised to unity today so that comoving scales become physical scales then. The Hubble parameter today is $H_0 \equiv 100\, h\, \text{km}/s /\Mpc$. For numerical estimates we use the $\Lambda$CDM model with parameters estimated from WMAP-7, ACT-2008, $H_0$ and BAO data given in table 5 of ref.\ \cite{Dunkley:2010ge}.  The total matter and baryon density fractions are $\Omega_m \equiv \rho_m/\rho_c$ and $\Omega_b \equiv \rho_b/\rho_c$, with the critical energy density $\rho_c\equiv 3H_0^2 \mpl^2$. We also use the conversion $1\,\Gauss\approx 4.8\times 10^{56}\,\Mpc^{-2}$ in Heaviside-Lorentz units.

One can introduce the helicity basis as
\be
  \Veps^\Vk_{\pm} \equiv \frac{1}{\sqrt{2}}\left( \Veps^\Vk_1 \pm i\Veps^\Vk_2 \right) \,,
\ee
where $\left(\Veps^\Vk_{1}\,,\,\Veps^\Vk_2\,,\,\hat\Vk\right)$ form an orthonormal spatial comoving basis with $|\Veps^\Vk_{i}|^2=1 \,,\ \hat\Vk=\Vk/k$. In the helicity basis, the vector potential takes the form $\cA_+\Veps_++\cA_-\Veps_-$. Thus, one can study the evolution of the Fourier modes, $\cA_h(\eta,k)$, with respect to the helicity basis for the polarisation states, $h=\pm$.

If the generated magnetic field is statistically homogeneous and isotropic, its spectrum is determined by two scalar functions $P_S(k)$ and $P_A(k)$. Due to the divergence-free nature of the magnetic fields, the two-point function for a comoving wave vector $\Vk$ in Fourier space can be written as
\be
  \langle \widetilde{B}_i(\eta,\Vk) \widetilde{B}^*_j(\eta,\Vq) \rangle
  \ =\ \frac{(2\pi)^3}{2} \delta(\Vk-\Vq)
  \Big\{(\delta_{ij} -\hat{k}_i\hat{k}_j) P_S(\eta,k) - i \epsilon_{ijn} \hat{k}_n P_A(\eta,k) \Big\}
\ee
where $P_S$ and $P_A$ are the symmetric and anti-symmetric parts of the power spectrum, respectively. The symmetric part of the spectrum determines the energy density:
\be
  \langle \widetilde{B}_i(\eta,\Vk) \widetilde{B}^*_i(\eta,\Vq) \rangle
  \ =\  (2\pi)^3\delta(\Vk-\Vq)P_S(\eta,k) \, .
\ee
With respect to the helicity basis the spectra can be directly written as
\be\label{e:PSA}
  P_{S/A}(\eta,k)\ =\ k^2\left( |\cA_+(\eta,k)|^2 \pm |\cA_-(\eta,k)|^2 \right) \,,
\ee
where the upper (lower) sign corresponds to $P_S (P_A)$, respectively.

\section{Fluid and current evolution in the covariant approach}\label{app:covariant}

Let us briefly summaries the notation of the 1+3 covariant Ehlers-Ellis formalism \cite{Ellis:1998ct, Tsagas:2007yx}. We will mostly follow the notation of ref.\ \cite{Clarkson:2010uz} who's appendix gives a concise and quite complete collection of definitions, relations and relevant equations. We give the covariant energy-momentum conservation equations and derive the general evolution equation of the bulk velocity field for a plasma without anisotropic stress in its rest frame.

\subsection{The covariant approach}

The covariant approach employs a time-like unit 4-velocity field, $u^\mu$ with $u^2\equiv g_\mn u^\mu u^\nu = -1$ to define a 1+3 splitting of space-time. The frame $u^\mu$ represents a family of comoving ``observers''. Then, all vectors and tensors are irreducibly decomposed w.r.t.\  $u^\mu$, the projection tensor $h^\mn\equiv g^\mn+u^\mu u^\nu$ and the projected alternating tensor $\veps_{\mn\alpha}\equiv\eta_{\mn\ab}u^\beta$ where $\eta_{\mn\ab}\equiv -\sqrt{-g}\delta^0_{[\mu}\delta^1_\nu \delta^2_\alpha \delta^3_{\beta]}$. The covariant derivative parallel and orthogonal to $u^\mu$ are denoted by a dot and $\D_\mu$ respectively. For a generic tensor $Y$ they read
\bea
\dot Y^{\mu\cdots}_{\phantom{\mu\cdots}\cdots\nu} &\equiv&
  u^\alpha\nabla_\alpha Y^{\mu\cdots}_{\phantom{\mu\cdots}\cdots\nu} \,,
\\
\D_\lambda Y^{\mu\cdots}_{\phantom{\mu\cdots}\cdots\nu} &\equiv&
  h_\lambda^\alpha h^\mu_\beta \cdots h_\nu^\gamma \nabla_\alpha
  Y^{\beta\cdots}_{\phantom{\beta\cdots}\cdots\gamma} \,.
\eea
The geometry as measured by the family of observers $u^\mu$ is described by the irreducible components of $\nabla_\nu u_\mu$
\be
\nabla_\nu u_\mu = -\dot u_\mu u_\nu +\frac{1}{3}\Theta h_\mn +\sigma_\mn +\omega_\mn\,.
\ee
These are the \emph{kinematic or geometric quantities} of $u^\mu$: the expansion $\Theta\equiv \D^\alpha u_\alpha$, the acceleration $\dot u_\mu \equiv u^\alpha\nabla_\alpha u_\mu$, the shear $\sigma_\mn\equiv \D_{\la\mu}u_{\nu\ra}$, and the vorticity $\omega_\mn\equiv \D_{[\mu}u_{\mu]}$. The vorticity vector is $\omega^\mu\equiv\veps^{\mu\ab}\omega_\ab=\veps^{\mu\ab}\D_\alpha u_\beta$. Here the projected symmetric trace-free parts are defined as
\be
V^{\la\mu\ra} \equiv h^\mu_\alpha V^\alpha \,,\quad
Y^{\la\mu\ra\nu} \equiv h^\mu_\alpha Y^{\alpha\nu} \,,\qquad
Y^{\la\mn\ra} \equiv \[h^{(\mu}_{\alpha\phantom{\beta}} h^{\nu)}_\beta
  -\frac{1}{3}h^\mn h_{\ab}\] Y^{\ab} \,,
\ee
and indices in round (square) brackets are being (anti-)symmetrised. Einstein's equations and the Bianchi identities then reduce to evolution equations and constraints for the geometric quantities, see references given above.

Next, we decompose the energy-momentum tensor of the species $s$, $T_s^\mn$, w.r.t.\ $u^\mu$ into its \emph{dynamic quantities} or \emph{fluid variables}: energy density $\rho_s\equiv u_\alpha u_\beta T_s^\ab$, isotropic pressure (incl.\ possible bulk viscosity) $p_s\equiv\frac{1}{3}h_\ab T_s^\ab$, energy flux (incl.\ possible heat flux) $q_s^\mu\equiv -u_\alpha T^{\la\mu\ra\alpha}_s$ and anisotropic stress (incl.\ possible shear viscosity) $\pi_s^\mn\equiv T^{\la\mn\ra}_s$. In a generic frame we then have
\be
T^\mn_s = \rho_s u^\mu u^\nu + p_s h^\mn + 2q_s^{(\mu} u^{\nu)} + \pi_s^\mn \, ,
\ee
even if the species is well described by a perfect fluid. The reason why $q_s^\mu$ and $\pi_s^\mn$ appear is that $u^\mu$ is not generally the rest frame of the species $s$.

\subsection{Observer frames}

Several choices of frame are possible and discussed in the literature. The physics do not depend on the choice of frame, but one frame may be more practical than the other. In the context of plasmas, two frames are interesting to note: the particle (or Eckart) frame in which the particle flux of the plasma vanishes, and the energy (or Landau) frame in which the energy flux vanishes. In the absence of heat flux, i.e.\ in thermal equilibrium, the two frames are equivalent. In this case we use the term \emph{rest frame} for the frame, $u_s^\mu$, defined by requiring $0=q_s^{*\mu} \equiv -u_{s\,\alpha} T^{\la\mu\ra\alpha}_s$. For more details on these frames see for instance \cite{Kandus:2007ap}.

Finally, the algebraic transformations of the dynamic quantities from one frame to another are known exactly, see e.g.\ the appendix of \cite{Clarkson:2010uz}. Here we only give the transformations from the rest frame to a generic frame, for a species without heat flux nor anisotropic stress in its energy frame. One decomposes the rest frame w.r.t.\ the generic frame according to
\be
u_s^\mu = \gamma_s\left(u^\mu + v^\mu_s\right) \quad,\quad u_\mu v^\mu_s = 0
\quad,\quad \gamma_s \equiv \left(1-v_s^2\right)^{-1/2} \, ,
\ee
where $\gamma_s$ is the local Lorentz factor of the transformation, and $v_s^\mu$ is the relative velocity. Then the dynamic quantities measured in the generic frame are
\be
\rho_s &=& \rho_s^* + \gamma_s^2 v_s^2(\rho_s^*+p_s^*) \,,
\label{trafo:rho}
\\
p_s &=& p_s^* + \tfrac{1}{3}\gamma_s^2 v_s^2(\rho_s^*+p_s^*) \,,
\label{trafo:p}
\\
q_s^\mu &=& \gamma_s^2(\rho_s^*+p_s^*)v_s^\mu \,,
\label{trafo:q}
\\
\pi_s^\mn &=& \gamma_s^2(\rho_s^*+p_s^*)v_s^{\la\mu}v_s^{\nu\ra} \,,
\label{trafo:pi}
\ee
and remember we assumed, $q_s^{*\mu}=0=\pi_s^{*\mn}$, in the rest frame. The other way around it can be seen that
\be
\rho_s^* &=& \rho_s - v_s^2\, \frac{\rho_s+p_s}{1+\frac{1}{3}v_s^2} \,,
\\
p_s^* &=& p_s - \frac{1}{3}v_s^2\, \frac{\rho_s+p_s}{1+\frac{1}{3}v_s^2} \,,
\label{eq:pstar}
\ee
and the equalities
\be
\gamma_s^2\(\rho_s^*+p_s^*\)\ =\ \rho_s+p_s^*\ =\ \frac{\rho_s+p_s}{1+\frac{1}{3}v_s^2} \,,
\ee
are useful as well.

\subsection{Energy-momentum conservation and the velocity evolution equation} \label{app:velocity}

The energy-momentum conservation for a certain species $s$ is written as $\nabla_\alpha T_s^{\mu\alpha}=\cC_s^\mu$. As a consequence of the Bianchi identities the total energy-momentum tensor is covariantly conserved and therefore $\sum_s\cC_s^\mu =0$. Projecting the conservation equation parallel and orthogonal to $u^\mu$ we have
\be
\hspace{-5mm}
\dot\rho_s + \Theta(\rho_s+p_s) + \Big(2\dot u_\alpha+\D_\alpha\Big) q_s^\alpha
  + \sigma_\ab\pi_s^{\ab} &=& \cC_s \,,
\label{eq:rhodot}
\\
\hspace{-5mm}
\dot q_s^{\la\mu\ra} + \frac{4}{3}\Theta q_s^\mu + \Big(\sigma^\mu_{\phantom{\mu}\alpha} 
  + \omega^\mu_{\phantom{\mu}\alpha}\Big) q_s^\alpha
  + (\rho_s+p_s)\dot u^\mu + \D^\mu p_s
  + \Big(\dot u_\alpha + \D_\alpha\Big) \pi_s^{\mu\alpha} &=& \cC_s^{\la\mu\ra} \,,
\label{eq:qdot}
\ee
where $\cC_s\equiv -u_\alpha \cC_s^\alpha$ and $\cC_s^{\la\mu\ra}\equiv h^\mu_\alpha \cC_s^\alpha$ are the energy and momentum transfer rates from all other species to species $s$. The momentum conservation (\ref{eq:qdot}) is an equation for the energy flux, $q_s^\mu$, measured in the generic frame $u^\mu$. To derive the evolution equation for the velocity, $v_s^\mu$, of the $s$ species w.r.t.\ $u^\mu$, we use the transformations (\ref{trafo:rho}) -- (\ref{trafo:pi}) and the energy conservation (\ref{eq:rhodot}). We notice that $q_s^\mu=(\rho_s+p_s^*)v_s^\mu$. Then we arrive at\footnote{Nearly the same expression was found in ref.\ \cite{Kandus:2007ap}, see their eq.\ (33). The differences are of order $v_s^2$ and are caused by incorrect frame transformations, their eqs.\ (11) and (12). Even though the terms add up to the correct energy-momentum tensor in their eq.\ (9), the anisotropic stress in their eq.\ (12) is not trace-free as it should. The trace needs to be subtracted and added to the pressure, see our (\ref{trafo:pi}) and \cite{Clarkson:2010uz}. The authors of ref.\ \cite{Kandus:2007ap} notice this in the text and carry on, but later do not realise that this leads to a certain confusion about the pressure when using the energy-momentum conservation equations, see their eqs.\ (25) and (26).}
\be
\dot{v}_s^{\la\mu\ra} &=& \[-\frac{1}{3}(1-v_s^2)\Theta 
  + \Big(\dot{u}_\alpha +\sigma_\ab v_s^\beta\Big)v_s^\alpha -v_s^\alpha\D_\alpha\] v_s^\mu
  - \Big[\sigma^\mu_{\phantom{\mu}\alpha} +\omega^\mu_{\phantom{\mu}\alpha}\Big]v_s^\alpha 
\nn \\
&& 
  - \dot{u}^\mu +\frac{1+\frac{1}{3}v_s^2}{\rho_s+p_s} \[ \cC_s^{\la\mu\ra}
  - \cC_s v_s^\mu - \D^\mu p_s^* - \dot{p}_s^* v_s^\mu \] \,.
\label{eq:vdot1}
\ee
Notice that $p_s^*$ is the pressure as measured in the rest frame of the $s$ species and is given in eq.\ (\ref{eq:pstar}) in terms of the dynamic quantities in the generic frame $u^\mu$. See also section \ref{app:presspert} on pressure perturbations.

\subsection{Current evolution equation} \label{app:current}

Again, we follow ref.\ \cite{Kandus:2007ap}. The particle number flux of each species $s$ in the generic frame $u^\mu$ decomposes as
\be
N_s^\mu = n_s \( u^\mu + v_s^\mu \)
\ee
with $n_s$ being the number density as seen in $u^\mu$. Thus, particles with charge $e_s$ give rise to a partial charge current of $j_s^\mu = e_s N_s^\mu$ that we decompose as follows $j_s^\mu =Q_s u^\mu + J_s^\mu$ with the partial charge density and spatial currents being
\be
Q_s = e_s n_s \,, \qquad  J_s^\mu =e_s n_s v_s^\mu \,.
\label{eq:partialcurrent}
\ee
The total charge density and spatial current are $Q=\sum_s Q_s$ and $J^\mu=\sum_s J_s^\mu$.

To get the evolution equation of the total current, we take the time derivative of the spatial current
\be
\dot{J}^{\la\mu\ra} = \sum_s e_s \( \dot{n}_s v_s^\mu + n_s \dot v_s^{\la\mu\ra} \) \,.
\label{eq:jdot1}
\ee
The evolution equation for the velocity is given above, but we also need an equation for the number density. Let us write the particle number conservation as $\nabla_\alpha N_s^\alpha = n_s\Delta_s$, where $\Delta_s$ is the rate of $s$ particles added or removed due to inelastic processes. By projecting the covariant derivative we find
\be
\dot{n}_s  = -\(\Theta + \dot{u}_\alpha v_s^\alpha +\D_\alpha v_s^\alpha - \Delta_s \)n_s
- v_s^\alpha\D_\alpha n_s \,.
\label{eq:ndot}
\ee
Finally, we insert eqs.\ (\ref{eq:ndot}) and (\ref{eq:vdot1}) into (\ref{eq:jdot1}) to find
\be
\dot{J}^{\la\mu\ra} &=& - \frac{4}{3}\Theta J^\mu
 - \Big[\sigma^\mu_{\phantom{\mu}\alpha}+\omega^\mu_{\phantom{\mu}\alpha}\Big]J^\alpha
 - Q \dot{u}^\mu
\nn \\ &&
 + \sum_s \Bigg\{ \[ \frac{1}{3}v_s^2\Theta +\sigma_\ab v_s^\alpha v_s^\beta \] J_s^\mu
 - \D_\alpha \big( v_s^\alpha J_s^\mu \big) 
\nn \\ &&
 - \frac{1+\frac{1}{3}v_s^2}{\rho_s+p_s}\[ e_s n_s \D^\mu p_s^* +\dot{p}_s^* J_s^\mu \]
\nn \\ &&
 + \frac{1+\frac{1}{3}v_s^2}{\rho_s+p_s}\[ e_s n_s \cC_s^{\la\mu\ra} - \cC_s J_s^\mu \]
 + \Delta_s J_s^\mu  \Bigg\} \,.
\label{eq:jdotfull}
\ee
The first line describes geometrical effects on the total current: the first and second terms are due to isotropic, shear and vortical expansion, while the third term is due to the local charge separation coupling to the acceleration of the frame $u^\mu$. The second line describes non-linear dynamical effects from the multi-component plasma. The third line describes the effects from pressure fluctuations that are known as the Biermann battery. The last line is due to the interactions of the plasma. The first and second term are the momentum and energy transfer between the species, respectively. They also contain the electromagnetic forces that give rise to the Hall effect and ultimately to the generalised Ohm law. The third term comes from particle creation and annihilation adding or removing charges.

\subsection{Pressure perturbations} \label{app:presspert}

To model the pressure in the rest frame we define two relevant quantities: the temporal and the spatial sound speeds, $c_t$ and $c_s$, respectively:
\be
u_s^\alpha\nabla_\alpha p_s \equiv c_{t,s}^2\, u_s^\alpha\nabla_\alpha \rho_s \,, \qquad
\D_s^\mu p_s^*=c_{s,s}^2\, \D_s^\mu \rho_s^* \,.
\ee
Here $u_s^\alpha\nabla_\alpha$ is the time derivative in the rest frame $u_s^\mu$ and $\D_s^\mu$ is the spatial derivative projected orthogonal to $u_s^\mu$. Both sound speeds describe how the pressure reacts to changes in the density. In addition, the temporal sound speed encodes possible changes in the pressure due to the time evolution of the entropy, while the spatial sound speed also contains the effect of spatial entropy fluctuations. In case of adiabatic time evolution, $c_t=c_a$, with $c_a$ being the standard adiabatic sound speed. In the absence of spatial entropy fluctuations, $c_s=c_a$, as well.

We are interested in the combination $\D^\mu p_s^* +\dot{p}_s^* v_s^\mu$ that appears in the velocity and current evolution equations above. We transform the derivatives, introduce the sound speeds and transform everything back into a generic frame $u^\mu$. We find
\be
\hspace{-6mm}
\D^\mu p_s^* +\dot{p}_s^* v_s^\mu &=& 
\frac{c_{s,s}^2}{1+c_{s,s}^2 v_s^2} \! \left\{
 \(1 - v_s^2\) \! \[ \D^\mu\rho_s + \dot\rho_s v_s^\mu \]
- \frac{\rho_s+p_s}{1+\frac{1}{3}v_s^2} \! \[ \D^\mu\big(v_s^2\big) 
  + \big(v_s^2\dot{\big)}v_s^\mu \] \right\}
\nn \\
&=& c_{s,s}^2 \Bigg\{ \D^\mu\rho_s -\! (\rho_s+p_s) \[ \Theta v_s^\mu 
  +  (\D_\alpha v_s^\alpha)v_s^\mu +\D^\mu \! \big(v_s^2\big)  \]
 + \cC_c v_s^\mu + \order{3} \Bigg\}
\ee
where, in the second step, we use the energy conservation equation (\ref{eq:rhodot}) and took along only terms up to second order in perturbations.

\bibliography{mfde_refs}

\providecommand{\href}[2]{#2}\begingroup\raggedright\begin{thebibliography}{10}

\bibitem{Kronberg:1993vk}
P.~P. Kronberg, {\it {Extragalactic magnetic fields}},  {\em Rept.Prog.Phys.}
  {\bf 57} (1994) 325--382.

\bibitem{Han:2002ns}
J.-L. Han and R.~Wielebinski, {\it {Milestones in the observations of cosmic
  magnetic fields}},  {\em Chin.J.Astron.Astrophys.} {\bf 2} (2002) 293--324,
  [\href{http://xxx.lanl.gov/abs/astro-ph/0209090}{{\tt astro-ph/0209090}}].

\bibitem{Govoni:2004as}
F.~Govoni and L.~Feretti, {\it {Magnetic field in clusters of galaxies}},  {\em
  Int.J.Mod.Phys.} {\bf D13} (2004) 1549--1594,
  [\href{http://xxx.lanl.gov/abs/astro-ph/0410182}{{\tt astro-ph/0410182}}].

\bibitem{Clarke:2000bz}
T.~Clarke, P.~Kronberg, and H.~Boehringer, {\it {A New radio - X-ray probe of
  galaxy cluster magnetic fields}},  {\em Astrophys.J.} {\bf 547} (2001)
  L111--L114, [\href{http://xxx.lanl.gov/abs/astro-ph/0011281}{{\tt
  astro-ph/0011281}}].

\bibitem{Tavecchio:2010mk}
F.~Tavecchio, G.~Ghisellini, L.~Foschini, G.~Bonnoli, G.~Ghirlanda, et~al.,
  {\it {The intergalactic magnetic field constrained by Fermi/LAT observations
  of the TeV blazar 1ES 0229+200}},  {\em Mon.Not.Roy.Astron.Soc.} {\bf 406}
  (2010) L70--L74, [\href{http://xxx.lanl.gov/abs/1004.1329}{{\tt
  arXiv:1004.1329}}].

\bibitem{Ando:2010rb}
S.~Ando and A.~Kusenko, {\it {Evidence for Gamma-Ray Halos Around Active
  Galactic Nuclei and the First Measurement of Intergalactic Magnetic Fields}},
   {\em Astrophys.J.} {\bf 722} (2010) L39,
  [\href{http://xxx.lanl.gov/abs/1005.1924}{{\tt arXiv:1005.1924}}].

\bibitem{Neronov:1900zz}
A.~Neronov and I.~Vovk, {\it {Evidence for strong extragalactic magnetic fields
  from Fermi observations of TeV blazars}},  {\em Science} {\bf 328} (2010)
  73--75, [\href{http://xxx.lanl.gov/abs/1006.3504}{{\tt arXiv:1006.3504}}].

\bibitem{Dolag:2010ni}
K.~Dolag, M.~Kachelriess, S.~Ostapchenko, and R.~Tomas, {\it {Lower limit on
  the strength and filling factor of extragalactic magnetic fields}},  {\em
  Astrophys.J.} {\bf 727} (2011) L4,
  [\href{http://xxx.lanl.gov/abs/1009.1782}{{\tt arXiv:1009.1782}}].

\bibitem{Essey:2010nd}
W.~Essey, S.~Ando, and A.~Kusenko, {\it {Determination of intergalactic
  magnetic fields from gamma ray data}},  {\em Astropart.Phys.} {\bf 35} (2011)
  135--139, [\href{http://xxx.lanl.gov/abs/1012.5313}{{\tt arXiv:1012.5313}}].

\bibitem{Saveliev:2012ea}
A.~Saveliev, K.~Jedamzik, and G.~Sigl, {\it {Time Evolution of the Large-Scale
  Tail of Non-Helical Primordial Magnetic Fields with Back-Reaction of the
  Turbulent Medium}},  {\em Phys.Rev.} {\bf D86} (2012) 103010,
  [\href{http://xxx.lanl.gov/abs/1208.0444}{{\tt arXiv:1208.0444}}].

\bibitem{Tevzadze:2012kk}
A.~G. Tevzadze, L.~Kisslinger, A.~Brandenburg, and T.~Kahniashvili, {\it
  {Magnetic Fields from QCD Phase Transitions}},  {\em Astrophys.J.} {\bf 759}
  (2012) 54, [\href{http://xxx.lanl.gov/abs/1207.0751}{{\tt arXiv:1207.0751}}].

\bibitem{Jedamzik:2010cy}
K.~Jedamzik and G.~Sigl, {\it {The Evolution of the Large-Scale Tail of
  Primordial Magnetic Fields}},  {\em Phys.Rev.} {\bf D83} (2011) 103005,
  [\href{http://xxx.lanl.gov/abs/1012.4794}{{\tt arXiv:1012.4794}}].

\bibitem{Caprini:2009pr}
C.~Caprini, R.~Durrer, and E.~Fenu, {\it {Can the observed large scale magnetic
  fields be seeded by helical primordial fields?}},  {\em JCAP} {\bf 0911}
  (2009) 001, [\href{http://xxx.lanl.gov/abs/0906.4976}{{\tt
  arXiv:0906.4976}}].

\bibitem{Urban:2009sw}
F.~R. Urban and A.~R. Zhitnitsky, {\it {Large-Scale Magnetic Fields, Dark
  Energy and QCD}},  {\em Phys.Rev.} {\bf D82} (2010) 043524,
  [\href{http://xxx.lanl.gov/abs/0912.3248}{{\tt arXiv:0912.3248}}].

\bibitem{Subramanian:2009fu}
K.~Subramanian, {\it {Magnetic fields in the early universe}},  {\em
  Astron.Nachr.} {\bf 331} (2010) 110--120,
  [\href{http://xxx.lanl.gov/abs/0911.4771}{{\tt arXiv:0911.4771}}].

\bibitem{Kandus:2010nw}
A.~Kandus, K.~E. Kunze, and C.~G. Tsagas, {\it {Primordial magnetogenesis}},
  {\em Phys.Rept.} {\bf 505} (2011) 1--58,
  [\href{http://xxx.lanl.gov/abs/1007.3891}{{\tt arXiv:1007.3891}}].

\bibitem{Widrow:2011hs}
L.~M. Widrow, D.~Ryu, D.~Schleicher, K.~Subramanian, C.~G. Tsagas, et~al., {\it
  {The First Magnetic Fields}},  {\em Space Sci.Rev.} {\bf 166} (2012) 37--70,
  [\href{http://xxx.lanl.gov/abs/1109.4052}{{\tt arXiv:1109.4052}}].

\bibitem{Ryu:2011hu}
D.~Ryu, D.~R. Schleicher, R.~A. Treumann, C.~G. Tsagas, and L.~M. Widrow, {\it
  {Magnetic fields in the Large-Scale Structure of the Universe}},  {\em Space
  Sci.Rev.} {\bf 166} (2012) 1--35,
  [\href{http://xxx.lanl.gov/abs/1109.4055}{{\tt arXiv:1109.4055}}].

\bibitem{Harrison:1970}
E.~R. Harrison, {\it {Generation of magnetic fields in the radiation era}},
  {\em Mon. Not. Roy. Astron. Soc.} {\bf 147} (1970) 279.

\bibitem{Harrison:1973zz}
E.~Harrison, {\it {Origin of Magnetic Fields in the Early Universe}},  {\em
  Phys.Rev.Lett.} {\bf 30} (1973) 188--190.

\bibitem{Hollenstein:2007kg}
L.~Hollenstein, C.~Caprini, R.~Crittenden, and R.~Maartens, {\it {Challenges
  for creating magnetic fields by cosmic defects}},  {\em Phys.Rev.} {\bf D77}
  (2008) 063517, [\href{http://xxx.lanl.gov/abs/0712.1667}{{\tt
  arXiv:0712.1667}}].

\bibitem{Berezhiani:2003ik}
Z.~Berezhiani and A.~D. Dolgov, {\it {Generation of large scale magnetic fields
  at recombination epoch}},  {\em Astropart.Phys.} {\bf 21} (2004) 59--69,
  [\href{http://xxx.lanl.gov/abs/astro-ph/0305595}{{\tt astro-ph/0305595}}].

\bibitem{Matarrese:2004kq}
S.~Matarrese, S.~Mollerach, A.~Notari, and A.~Riotto, {\it {Large-scale
  magnetic fields from density perturbations}},  {\em Phys.Rev.} {\bf D71}
  (2005) 043502, [\href{http://xxx.lanl.gov/abs/astro-ph/0410687}{{\tt
  astro-ph/0410687}}].

\bibitem{Gopal:2004ut}
R.~Gopal and S.~Sethi, {\it {Generation of magnetic field in the
  pre-recombination era}},  {\em Mon.Not.Roy.Astron.Soc.} {\bf 363} (2005)
  521--528, [\href{http://xxx.lanl.gov/abs/astro-ph/0411170}{{\tt
  astro-ph/0411170}}].

\bibitem{Ichiki:2007hu}
K.~Ichiki, K.~Takahashi, N.~Sugiyama, H.~Hanayama, and H.~Ohno, {\it {Magnetic
  Field Spectrum at Cosmological Recombination}},
  \href{http://xxx.lanl.gov/abs/astro-ph/0701329}{{\tt astro-ph/0701329}}.

\bibitem{Kobayashi:2007wd}
T.~Kobayashi, R.~Maartens, T.~Shiromizu, and K.~Takahashi, {\it {Cosmological
  magnetic fields from nonlinear effects}},  {\em Phys.Rev.} {\bf D75} (2007)
  103501, [\href{http://xxx.lanl.gov/abs/astro-ph/0701596}{{\tt
  astro-ph/0701596}}].

\bibitem{Takahashi:2007ds}
K.~Takahashi, K.~Ichiki, and N.~Sugiyama, {\it {Electromagnetic Properties of
  the Early Universe}},  {\em Phys.Rev.} {\bf D77} (2008) 124028,
  [\href{http://xxx.lanl.gov/abs/0710.4620}{{\tt arXiv:0710.4620}}].

\bibitem{Maeda:2008dv}
S.~Maeda, S.~Kitagawa, T.~Kobayashi, and T.~Shiromizu, {\it {Primordial
  magnetic fields from second-order cosmological perturbations:Tight coupling
  approximation}},  {\em Class.Quant.Grav.} {\bf 26} (2009) 135014,
  [\href{http://xxx.lanl.gov/abs/0805.0169}{{\tt arXiv:0805.0169}}].

\bibitem{Fenu:2010kh}
E.~Fenu, C.~Pitrou, and R.~Maartens, {\it {The seed magnetic field generated
  during recombination}},  {\em Mon.Not.Roy.Astron.Soc.} {\bf 414} (2011)
  2354--2366, [\href{http://xxx.lanl.gov/abs/1012.2958}{{\tt
  arXiv:1012.2958}}].

\bibitem{Maeda:2011uq}
S.~Maeda, K.~Takahashi, and K.~Ichiki, {\it {Primordial magnetic fields
  generated by the non-adiabatic fluctuations at pre-recombination era}},  {\em
  JCAP} {\bf 1111} (2011) 045, [\href{http://xxx.lanl.gov/abs/1109.0691}{{\tt
  arXiv:1109.0691}}].

\bibitem{Giovannini:2011tj}
M.~Giovannini, {\it {Reynolds numbers in the early Universe}},  {\em
  Phys.Lett.} {\bf B711} (2012) 327--331,
  [\href{http://xxx.lanl.gov/abs/1111.3867}{{\tt arXiv:1111.3867}}].

\bibitem{Lee:2001hj}
D.-S. Lee, W.-l. Lee, and K.-W. Ng, {\it {Primordial magnetic fields from dark
  energy}},  {\em Phys.Lett.} {\bf B542} (2002) 1--7,
  [\href{http://xxx.lanl.gov/abs/astro-ph/0109184}{{\tt astro-ph/0109184}}].

\bibitem{BeltranJimenez:2010uh}
J.~Beltran~Jimenez and A.~L. Maroto, {\it {Dark energy, non-minimal couplings
  and the origin of cosmic magnetic fields}},  {\em JCAP} {\bf 1012} (2010)
  025, [\href{http://xxx.lanl.gov/abs/1010.4513}{{\tt arXiv:1010.4513}}].

\bibitem{Kandus:2007ap}
A.~Kandus and C.~G. Tsagas, {\it {Generalized Ohm's law for relativistic
  plasmas}},  {\em Mon.Not.Roy.Astron.Soc.} {\bf 385} (2008) 883--892,
  [\href{http://xxx.lanl.gov/abs/0711.3573}{{\tt arXiv:0711.3573}}].

\bibitem{Maartens:1998xg}
R.~Maartens, T.~Gebbie, and G.~F. Ellis, {\it {Covariant cosmic microwave
  background anisotropies. 2. Nonlinear dynamics}},  {\em Phys.Rev.} {\bf D59}
  (1999) 083506, [\href{http://xxx.lanl.gov/abs/astro-ph/9808163}{{\tt
  astro-ph/9808163}}].

\bibitem{Uzan:1998mc}
J.-P. Uzan, {\it {Dynamics of relativistic interacting gases: From a kinetic to
  a fluid description}},  {\em Class.Quant.Grav.} {\bf 15} (1998) 1063--1088,
  [\href{http://xxx.lanl.gov/abs/gr-qc/9801108}{{\tt gr-qc/9801108}}].

\bibitem{Huba:2011}
J.~D. Huba, {\em {NRL Plasma Formulary}}.
\newblock National Research Laboratory, Washington DC, 2011.
\newblock
  \texttt{\ifpdf\href{http://wwwppd.nrl.navy.mil/nrlformulary/}{http://wwwppd.nrl.navy.mil/nrlformulary}\else{http://wwwppd.nrl.navy.mil/nrlformulary}\fi}.

\bibitem{Mukhanov:2005sc}
V.~Mukhanov, {\em {Physical foundations of cosmology}}.
\newblock Cambridge University Press, Cambridge, UK, 2005.

\bibitem{Ma:1995ey}
C.-P. Ma and E.~Bertschinger, {\it {Cosmological perturbation theory in the
  synchronous and conformal Newtonian gauges}},  {\em Astrophys.J.} {\bf 455}
  (1995) 7--25, [\href{http://xxx.lanl.gov/abs/astro-ph/9506072}{{\tt
  astro-ph/9506072}}].

\bibitem{Mongwane:2012gg}
B.~Mongwane, P.~K. Dunsby, and B.~Osano, {\it {Cosmic Electromagnetic Fields
  due to Perturbations in the Gravitational Field}},  {\em Phys.Rev.} {\bf D86}
  (2012) 083533, [\href{http://xxx.lanl.gov/abs/1203.6032}{{\tt
  arXiv:1203.6032}}].

\bibitem{Grasso:2000wj}
D.~Grasso and H.~R. Rubinstein, {\it {Magnetic fields in the early universe}},
  {\em Phys.Rept.} {\bf 348} (2001) 163--266,
  [\href{http://xxx.lanl.gov/abs/astro-ph/0009061}{{\tt astro-ph/0009061}}].

\bibitem{Giovannini:2003yn}
M.~Giovannini, {\it {The Magnetized universe}},  {\em Int.J.Mod.Phys.} {\bf
  D13} (2004) 391--502, [\href{http://xxx.lanl.gov/abs/astro-ph/0312614}{{\tt
  astro-ph/0312614}}].

\bibitem{Thompson:2012pj}
R.~I. Thompson, C.~Martins, and P.~Vielzeuf, {\it {Constraining cosmologies
  with fundamental constants I. Quintessence and K-Essence}},  {\em
  Mon.Not.Roy.Astron.Soc.} {\bf 428} (2013) 2232,
  [\href{http://xxx.lanl.gov/abs/1210.3031}{{\tt arXiv:1210.3031}}].

\bibitem{Avelino:2006gc}
P.~Avelino, C.~Martins, N.~J. Nunes, and K.~Olive, {\it {Reconstructing the
  dark energy equation of state with varying couplings}},  {\em Phys.Rev.} {\bf
  D74} (2006) 083508, [\href{http://xxx.lanl.gov/abs/astro-ph/0605690}{{\tt
  astro-ph/0605690}}].

\bibitem{Uzan:2010pm}
J.-P. Uzan, {\it {Varying Constants, Gravitation and Cosmology}},  {\em Living
  Rev.Rel.} {\bf 14} (2011) 2, [\href{http://xxx.lanl.gov/abs/1009.5514}{{\tt
  arXiv:1009.5514}}].

\bibitem{Ni:2011ti}
W.-T. Ni, A.~Balakin, and H.-H. Mei, {\it {Pseudoscalar-photon Interactions,
  Axions, Non-Minimal Extensions, and Their Empirical Constraints from
  Observations}},  in {\em {Conference in Honor of Murray Gell-Mann's 80th
  Birthday: Quantum Mechanics, Elementary Particles, Quantum Cosmology \&
  Complexity}}, pp.~526--535, 2011.
\newblock \href{http://xxx.lanl.gov/abs/1109.0581}{{\tt arXiv:1109.0581}}.

\bibitem{Urban:2010wa}
F.~R. Urban and A.~R. Zhitnitsky, {\it {The Parity Odd Universe, Dark Energy
  and QCD}},  {\em Phys.Rev.} {\bf D83} (2011) 123532,
  [\href{http://xxx.lanl.gov/abs/1011.2425}{{\tt arXiv:1011.2425}}].

\bibitem{Payez:2008pm}
A.~Payez, J.~Cudell, and D.~Hutsemekers, {\it {Axions and polarisation of
  quasars}},  {\em AIP Conf.Proc.} {\bf 1038} (2008) 211--219,
  [\href{http://xxx.lanl.gov/abs/0805.3946}{{\tt arXiv:0805.3946}}].

\bibitem{Martin:2007ue}
J.~Martin and J.~Yokoyama, {\it {Generation of Large-Scale Magnetic Fields in
  Single-Field Inflation}},  {\em JCAP} {\bf 0801} (2008) 025,
  [\href{http://xxx.lanl.gov/abs/0711.4307}{{\tt arXiv:0711.4307}}].

\bibitem{Demozzi:2009fu}
V.~Demozzi, V.~Mukhanov, and H.~Rubinstein, {\it {Magnetic fields from
  inflation?}},  {\em JCAP} {\bf 0908} (2009) 025,
  [\href{http://xxx.lanl.gov/abs/0907.1030}{{\tt arXiv:0907.1030}}].

\bibitem{Anber:2006xt}
M.~M. Anber and L.~Sorbo, {\it {N-flationary magnetic fields}},  {\em JCAP}
  {\bf 0610} (2006) 018, [\href{http://xxx.lanl.gov/abs/astro-ph/0606534}{{\tt
  astro-ph/0606534}}].

\bibitem{Durrer:2010mq}
R.~Durrer, L.~Hollenstein, and R.~K. Jain, {\it {Can slow roll inflation induce
  relevant helical magnetic fields?}},  {\em JCAP} {\bf 1103} (2011) 037,
  [\href{http://xxx.lanl.gov/abs/1005.5322}{{\tt arXiv:1005.5322}}].

\bibitem{Barnaby:2010vf}
N.~Barnaby and M.~Peloso, {\it {Large Nongaussianity in Axion Inflation}},
  {\em Phys.Rev.Lett.} {\bf 106} (2011) 181301,
  [\href{http://xxx.lanl.gov/abs/1011.1500}{{\tt arXiv:1011.1500}}].

\bibitem{Byrnes:2011aa}
C.~T. Byrnes, L.~Hollenstein, R.~K. Jain, and F.~R. Urban, {\it {Resonant
  magnetic fields from inflation}},  {\em JCAP} {\bf 1203} (2012) 009,
  [\href{http://xxx.lanl.gov/abs/1111.2030}{{\tt arXiv:1111.2030}}].

\bibitem{Ni:2007ar}
W.-T. Ni, {\it {From Equivalence Principles to Cosmology: Cosmic Polarization
  Rotation, CMB Observation, Neutrino Number Asymmetry, Lorentz Invariance and
  CPT}},  {\em Prog.Theor.Phys.Suppl.} {\bf 172} (2008) 49--60,
  [\href{http://xxx.lanl.gov/abs/0712.4082}{{\tt arXiv:0712.4082}}].

\bibitem{Alighieri:2010pu}
S.~d.~S. Alighieri, {\it {Cosmological Birefringence: an Astrophysical test of
  Fundamental Physics}},  \href{http://xxx.lanl.gov/abs/1011.4865}{{\tt
  arXiv:1011.4865}}.

\bibitem{Gruppuso:2011ci}
A.~Gruppuso, P.~Natoli, N.~Mandolesi, A.~De~Rosa, F.~Finelli, et~al., {\it
  {WMAP 7 year constraints on CPT violation from large angle CMB
  anisotropies}},  {\em JCAP} {\bf 1202} (2012) 023,
  [\href{http://xxx.lanl.gov/abs/1107.5548}{{\tt arXiv:1107.5548}}].

\bibitem{Alighieri:2010eu}
S.~d.~S. Alighieri, F.~Finelli, and M.~Galaverni, {\it {Limits on Cosmological
  Birefringence from the UV Polarization of Distant Radio Galaxies}},  {\em
  Astrophys.J.} {\bf 715} (2010) 33--38,
  [\href{http://xxx.lanl.gov/abs/1003.4823}{{\tt arXiv:1003.4823}}].

\bibitem{Goldhaber:2008xy}
A.~S. Goldhaber and M.~M. Nieto, {\it {Photon and Graviton Mass Limits}},  {\em
  Rev.Mod.Phys.} {\bf 82} (2010) 939--979,
  [\href{http://xxx.lanl.gov/abs/0809.1003}{{\tt arXiv:0809.1003}}].

\bibitem{Ellis:1998ct}
G.~F. Ellis and H.~van Elst, {\it {Cosmological models: Cargese lectures
  1998}},  {\em NATO Adv.Study Inst.Ser.C.Math.Phys.Sci.} {\bf 541} (1999)
  1--116, [\href{http://xxx.lanl.gov/abs/gr-qc/9812046}{{\tt gr-qc/9812046}}].

\bibitem{Tsagas:2007yx}
C.~G. Tsagas, A.~Challinor, and R.~Maartens, {\it {Relativistic cosmology and
  large-scale structure}},  {\em Phys.Rept.} {\bf 465} (2008) 61--147,
  [\href{http://xxx.lanl.gov/abs/0705.4397}{{\tt arXiv:0705.4397}}].

\bibitem{Clarkson:2010uz}
C.~Clarkson and R.~Maartens, {\it {Inhomogeneity and the foundations of
  concordance cosmology}},  {\em Class.Quant.Grav.} {\bf 27} (2010) 124008,
  [\href{http://xxx.lanl.gov/abs/1005.2165}{{\tt arXiv:1005.2165}}].

\bibitem{Dunkley:2010ge}
J.~Dunkley, R.~Hlozek, J.~Sievers, V.~Acquaviva, P.~Ade, et~al., {\it {The
  Atacama Cosmology Telescope: Cosmological Parameters from the 2008 Power
  Spectra}},  {\em Astrophys.J.} {\bf 739} (2011) 52,
  [\href{http://xxx.lanl.gov/abs/1009.0866}{{\tt arXiv:1009.0866}}].

\end{thebibliography}\endgroup
\bibliographystyle{JHEP}
\end{document}